%% file: main.tex
\documentclass[acmlarge]{acmart}
\renewcommand\footnotetextcopyrightpermission[1]{} 

\usepackage{subcaption}
\usepackage[ruled,vlined]{algorithm2e}

\usepackage{enumitem}
\usepackage{tikz}
\usepackage{siunitx}
\newcommand*\circled[1]{\tikz[baseline=(char.base)]{
            \node[shape=circle,draw,inner sep=2pt] (char) {#1};}}
            
\usepackage{xcolor}

\AtBeginDocument{%
  \providecommand\BibTeX{{%
    \normalfont B\kern-0.5em{\scshape i\kern-0.25em b}\kern-0.8em\TeX}}}

\begin{document}
\fancyfoot{}
\acmDOI{}
\title{Parallel Simulation of Quantum Networks with Distributed Quantum State Management}


\author{Xiaoliang Wu}
\affiliation{%
  \institution{Illinois Institute of Technology}
  \city{Chicago, IL}
  \country{United States}}
\email{xwu64@hawk.iit.edu}

\author{Alexander Kolar}
\affiliation{%
  \institution{University of Chicago}
  \city{Chicago, IL}
  \country{United States}}
\email{atkolar@uchicago.edu}

\author{Joaquin Chung}
\affiliation{%
  \institution{Argonne National Laboratory}
  \city{Lemont, IL}
  \country{United States}}
\email{chungmiranda@anl.gov}

\author{Dong Jin}
\affiliation{%
  \institution{Illinois Institute of Technology}
  \city{Chicago, IL}
  \country{}}
\affiliation{%
  \institution{University of Arkansas}
  \city{Fayetteville, AR}
\country{United States}}

\email{dongjin@uark.edu}

\author{Rajkumar Kettimuthu}
\affiliation{%
  \institution{Argonne National Laboratory}
  \city{Lemont, IL}
  \country{United States}}
\email{kettimut@anl.gov}

\author{Martin Suchara}
\affiliation{%
  \institution{Argonne National Laboratory}
  \city{Lemont, IL}
  \country{United States}}
\email{msuchara@anl.gov}

\newcommand{\todo}{{\color{red} \{...\} }}

\begin{abstract}
  Quantum network simulators offer the opportunity to cost-efficiently investigate potential avenues to building networks that scale with the number of users, communication distance, and application demands by simulating alternative hardware designs and control protocols. Several quantum network simulators have been recently developed with these goals in mind. However, as the size of the simulated networks increases, sequential execution becomes time consuming. Parallel execution presents a suitable method for scalable simulations of large-scale quantum networks, but the unique attributes of quantum information create some unexpected challenges. In this work we identify requirements for parallel simulation of quantum networks and develop the first parallel discrete event quantum network simulator by modifying the existing serial SeQUeNCe simulator. Our contributions include the design and development of a quantum state manager (QSM) that maintains shared quantum information distributed across multiple processes. We also optimize our parallel code by minimizing the overhead of the QSM and decreasing the amount of synchronization among processes. Using these techniques, we observe a speedup of 2 to 25 times when simulating a 1,024-node linear network with 2 to 128 processes. We also observe efficiency greater than 0.5 for up to 32 processes in a linear network topology of the same size and with the same workload. We repeat this evaluation with a randomized workload on a caveman network. Finally, we also introduce several methods for partitioning networks by mapping them to different parallel simulation processes. We released the parallel SeQUeNCe simulator as an open-source tool alongside the existing sequential version.
\end{abstract}




\maketitle
\thispagestyle{empty}

\input{intro}
\input{background}
\input{design}
\input{implementation}

\input{lookahead}
\input{evaluation}
\input{related_work}
\input{conclusion}

\begin{acks}
We thank Siu Man Chan for contributions to the SeQUeNCe software package and Gail Pieper for help with editing. The work of M. S., R. K. and A. K. is supported by the U.S. Department of Energy, Office of Science, National Quantum Information Science Research Centers. The work of X. W. and D. J. is partly sponsored by the Air Force Office of Scientific Research (AFOSR) under Grant YIP FA9550-17-1-0240 and the National Science Foundation (NSF) under Grant CNS-1730488 and DMR-1747426. We gratefully acknowledge the computing resources provided on Bebop, a high-performance computing cluster operated by the Laboratory Computing Resource Center at Argonne National Laboratory.
\end{acks}

\bibliographystyle{ACM-Reference-Format}
\bibliography{mybib}


\end{document}

%% file: intro.tex
\section{Introduction}

Over the last decade, the development of quantum networking testbeds has generated a lot of interest in the scientific community~\cite{Simon2017, darpa, ChinaNetwork, fermin-anl-link, fqnet, ukqnet}.
However, the scale of these demonstrators is limited to a handful of nodes and communication distances that typically do not exceed metropolitan scale. To advance the state of the art, simulations of quantum networks can help researchers develop and test new control protocols, plan and validate experiments, and compare alternative network architecture designs. They provide a cost-effective way to complement experimental research that develops real-world testbeds and prototypes.

A quantum network simulator must offer realism (accurate timing and representation of quantum states), flexibility (support of new types of hardware and protocols), and scalability (large number of nodes and events) for experimenters to evaluate their design and applications. Existing quantum network simulators such as SeQUeNCe~\cite{sequence}, Netsquid~\cite{netsquid}, and QuISP~\cite{matsuo2019quantum} achieve realism and flexibility by modeling quantum networks using sequential discrete-event simulation (DES). However, the sequential execution limits scalability since the simulation speed is limited by the performance of a single core. Simulating large-scale quantum networks through sequential execution takes a long time because quantum networks are more complex that their classical counterparts. The complexity of quantum network simulations comes from the need to model the behavior of several new types of hardware, and take into consideration processing and maintenance of quantum states, which typically generates a massive number of events. For example, the SeQUeNCe quantum network simulator tracks millions of events per second of simulation time, and one example of a simple metropolitan network simulation generated approximately 2 billion events~\cite{sequence}. Nevertheless, high performance of a quantum network simulator is necessary to allow researchers to design quantum networks and uncover potential risks before deployment in hardware testbeds.

Parallel discrete-event simulation (PDES)~\cite{pdes} improves the scalability of DES and significantly reduces the execution time by utilizing multiple processors to simultaneously execute events that ultimately advance the simulation of experiments faster than sequential simulation does. A quantum network (that is being simulated) could be partitioned into multiple subnetworks, each handled by a separate processor. The PDES synchronizes the states of subnetworks across different processes and advances the global simulation timeline. This technique has been extensively applied in PDES for classical networks~\cite{billion-nodes-ns3,ns3-perf} .
\textit{The key to successfully parallelizing simulations of quantum networks is the efficient synchronization of events and states among processes.}

In this work we parallelize SeQUeNCe using multiprocessing with an effective synchronization algorithm by considering specific quantum network characteristics. First, the Message Passing Interface (MPI)~\cite{mpi} is used to synchronize cross-process quantum channels and classical channels. Although multithreading can help speed up the execution on a single node, too many threads can lead to diminishing returns (e.g., when the number of threads exceeds the number of cores). MPI-based multiprocessing supports the execution of the simulation on a cluster of MPI nodes in both shared- and distributed-memory systems for scalability enhancement. Second, we design a \emph{hierarchical} quantum state manager (QSM) to store and synchronize the quantum state of simulated qubits. The QSM consists of one local QSM per process and a global QSM for handling cross-process entangled qubits. Processes use TCP-based socket communication to access the quantum states on the global QSM. Third, we develop a conservative synchronization algorithm to orchestrate MPI processes and QSM for parallel simulation. The algorithm enables MPI processes to exchange cross-process events at a dynamically computed synchronization point.

The design of our parallel quantum network simulator carefully considers requirements for scalability, accuracy, and usability. The \textbf{scalability} of the parallel simulator is determined by the performance of state synchronization among processes. We improve the simulation scalability in two ways. First, we save the communication time between the local QSMs and the global QSM by reducing the number of packets. The global QSM offloads portion of the computational work to the local QSMs and batches the request processing. Second, we reduce the overhead caused by an unbalanced workload by reducing the number of synchronization events. In particular, we use high lookahead values (within the constraints of conservative synchronization) based on the quantum network characteristics. 

The \textbf{accuracy} of a parallel simulator requires that the parallel execution of events does not affect the simulation results, in other  words, that the parallel execution is identical to the sequential execution. We design the synchronization algorithm by considering the communication between the local QSMs and the global QSM. The optimized lookahead value guarantees that the parallel event execution does not affect the causality of events. Additionally, we modify the functions to access the random number generator used in the simulation. Without this modification, events in PDES may trigger different actions because of different random numbers. These modified functions ensure that the sequence of random numbers used in PDES is identical to that in the sequential simulation. As a result, the parallelized simulator can correctly reproduce sequential results.

To improve \textbf{usability}, the kernel of the parallel simulator enables systematic model parallelization and hides functions such as QSM and random number generation from the simulation design. Notably, users can develop and use the same simulation setup for both parallel and sequential simulation.

We perform extensive evaluation of the parallelized SeQUeNCe with three network topologies:  linear topology, caveman graph~\cite{caveman}, and Internet autonomous system~\cite{as_net}. We simulate quantum networks consisting of 1,024 quantum routers with 1 to 128 processes using up to 4 nodes on the Bebop supercomputer~\cite{bebop}, a high-performance computing cluster located at Argonne National Laboratory. We use the speedup and execution time to evaluate the scalability of our simulator. The results show that PDES can speed up the networks with linear topology up to 25 times. We also analyze the sources of parallelization overhead. The overhead from the unbalanced workloads increases with the growing number of processes. Moreover, we study the performance improvement by reducing the QSM communication time (i.e., up to 11.5 times speedup) and enlarging the lookahead value (i.e., more than 50\% synchronization time reduction).

We summarize the contributions of this work as follows.

\begin{enumerate}
    \item We analyzed the requirements for parallel simulation and \textbf{designed a PDES-based simulation framework of quantum networks based on SeQUeNCe} that allow users to keep using the models and setup they already defined for the sequential version.
    \item We \textbf{developed a quantum state manager} that stores and synchronizes the state of simulated qubits across multiple processes.
    \item We conducted extensive performance evaluation of our parallel quantum network simulation with three network topologies on an HPC cluster, showing that \textbf{our approach provides up to 25$\times$ speed-up}.
\end{enumerate}

The remainder of the paper is organized as follows. In Section~\ref{sec:bg} we introduce the background on quantum information and quantum networks as well as the SeQUeNCe simulator and parallel simulation of classical networks. In Section~\ref{sec:design} we analyze the requirements for parallel simulation of quantum networks and present the design architecture of our simulator. In Section~\ref{sec:implementation} we describe the QSM design that maintains quantum states in parallel simulations. In Section~\ref{sec:async} we discuss lookahead optimization for reducing the simulation overhead. In Section~\ref{sec:eval} we evaluate the performance of the parallel simulator. In Section~\ref{sec:related} we describe related work. In Section~\ref{sec:conclusion} we conclude and briefly outline future work.

%% file: background.tex
\section{Background}
\label{sec:bg}

In this section we introduce a few elementary concepts in quantum information science. We also introduce the SeQUeNCe quantum network simulator and discuss the role of parallelization in classical network simulations.

\subsection{Quantum Information and Quantum Network Protocols}
\label{subsec:bg:qs}

\emph{Superposition} allows a quantum system to be simultaneously in two different states. These states can be mathematically represented as a probability distribution of the possible measurement outcomes on the quantum system. For example, the measurement of a particle in the superposition state $ | \psi \rangle := \alpha | 0 \rangle + \beta | 1 \rangle$ yields the measurement outcome $|0 \rangle$ with probability $|\alpha|^2$ and $|1 \rangle$ with probability $|\beta|^2$, where $\alpha$ and $\beta$ are two complex numbers. Upon measurement (on the z-axis), the quantum state $ | \psi \rangle $ collapses to either $|0\rangle$ or $|1\rangle$. Optical quantum networks use photons to carry quantum states. 

Multi-particle quantum systems can exhibit \emph{entanglement}, where the quantum state of each particle in the group cannot be described independently of the others. The simplest example is the EPR pair named after Einstein, Podolsky and Rosen~\cite{einstein1935EPR}: $| \phi^+ \rangle := \frac{1}{\sqrt{2}} (| 00 \rangle + | 11 \rangle)$. Here the measurement outcomes of the two particles are perfectly correlated with a 50\% chance of being $|0\rangle$, $|0\rangle$ and 50\% chance being $|1\rangle$, $|1\rangle$. The measurement of the first particle immediately collapses the state of the other particle to the same state, even if the particles are spatially separated. We note that with increasing number of particles, the space required to represent quantum states grows exponentially, presenting a significant challenge for computer simulations.

\begin{figure}[h]
  \centering
  \includegraphics[width=0.6\linewidth]{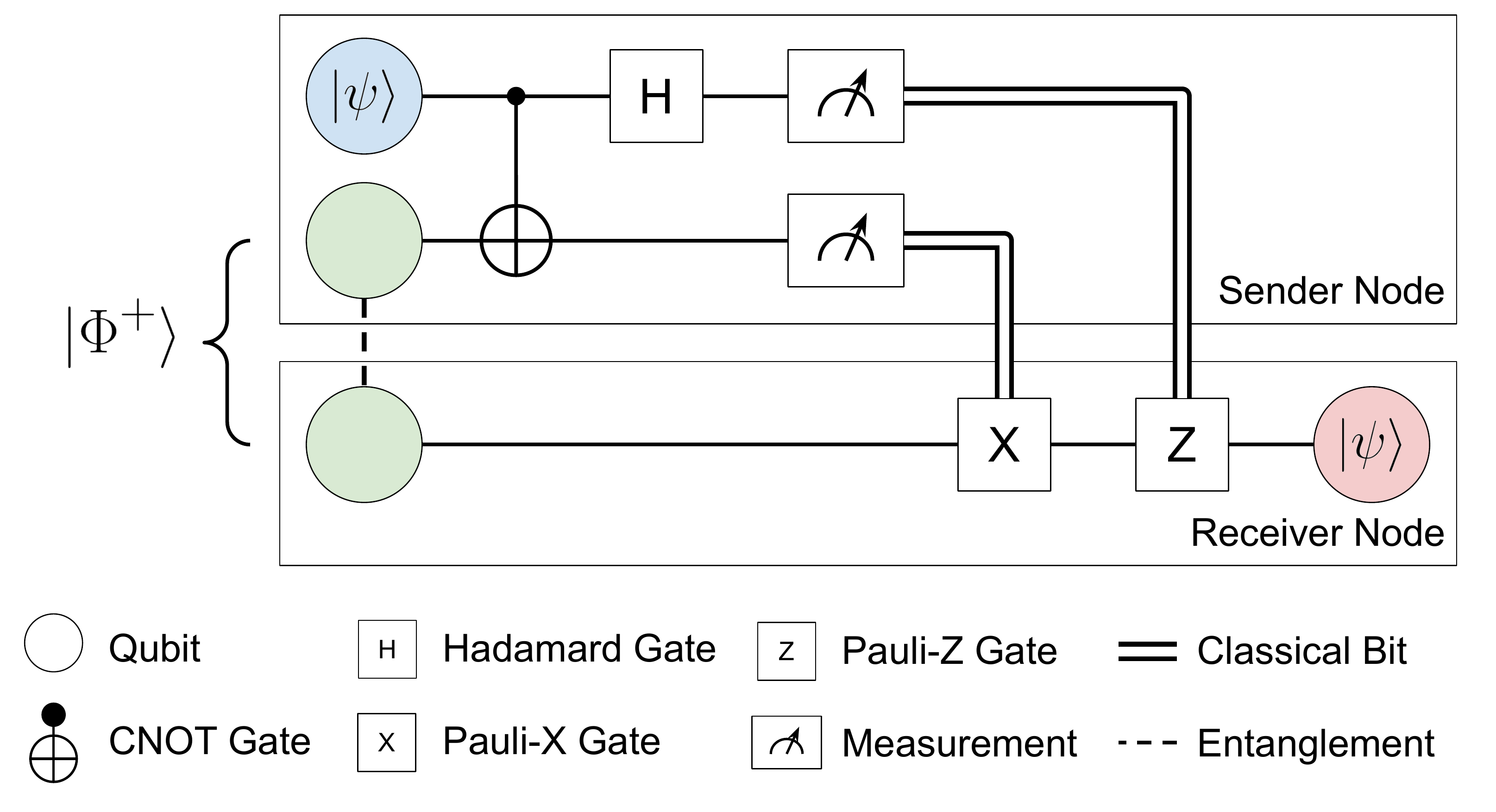}
  \caption{Quantum circuit for quantum teleportation.}
  \label{fig:circ_telep}
\end{figure}

One can use entanglement to \emph{teleport} an arbitrary quantum state over a long distance without worrying about state loss due to channel attenuation~\cite{Teleportation}. The teleportation protocol is depicted in Figure~\ref{fig:circ_telep}. Solid lines represent qubits, double lines classical bits, circles quantum states, and boxes quantum operations (quantum gates). Initially the sender and the receiver share an entangled qubit pair (green circles) that is distributed between them using the quantum network. The arbitrary quantum state $|\psi\rangle$ we want to transmit is prepared in another qubit (blue circle). The sender performs the indicated elementary operations on its two qubits (CNOT and Hadamard gates) and measures them. One of four possible measurement outcomes can be realized: 00, 01, 10, or 11. The measurement outcome is then sent to the receiver using a classical channel, and the receiver uses it to perform controlled quantum operations on its qubit in a process called Pauli frame correction. Upon the correction, it can be shown that the receiver's qubit (red circle) will be in state $|\psi\rangle$. Teleportation is one example of a quantum network protocol and it allows quantum state transmission by only distributing EPR pairs and classical information.

\begin{figure}[h]
  \centering
  \includegraphics[width=0.6\linewidth]{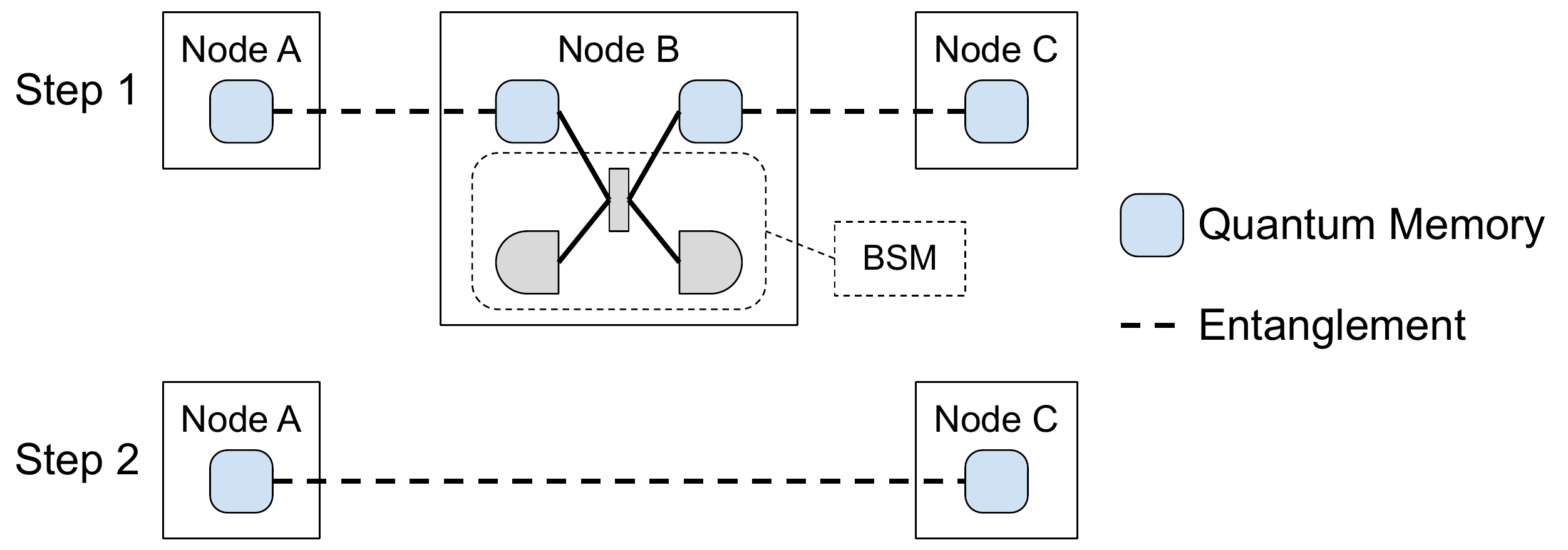}
  \caption{Two short-distance entangled pairs are used to create a long-distance entanglement with the swapping protocol.}
  \Description{Example of swapping protocol.}
  \label{fig:es}
\end{figure}

Reliable, long-distance entanglement distribution is an important service provided by quantum networks and many other protocols depend on it, including teleportation. The success probability of entanglement distribution decreases exponentially with distance. Fortunately, the \emph{entanglement swapping} protocol~\cite{swapping} can be used to ``stitch together'' multiple shorter-distance entanglements. The swapping protocol is illustrated in Figure~\ref{fig:es}. In the first step, entanglement is generated between the adjacent nodes A-B and B-C. Entangled photons may be stored in quantum memories to ensure that they are available precisely when needed~\cite{generation_schemes}. To create entanglement between nodes A-C, node B performs Bell state measurement (BSM) on its two qubits. The measurement outcome is sent from node B to either node A or node C to perform Pauli frame correction. After the correction the qubits on nodes A and C become entangled in the EPR state. This article simulates entanglement distribution and entanglement swapping in networks with up to 1,024 nodes, a computationally challenging problem.

\subsection{SeQUeNCe: A Customizable Quantum Network Simulator}
\begin{figure}[h]
  \centering
  \includegraphics[width=0.4\linewidth]{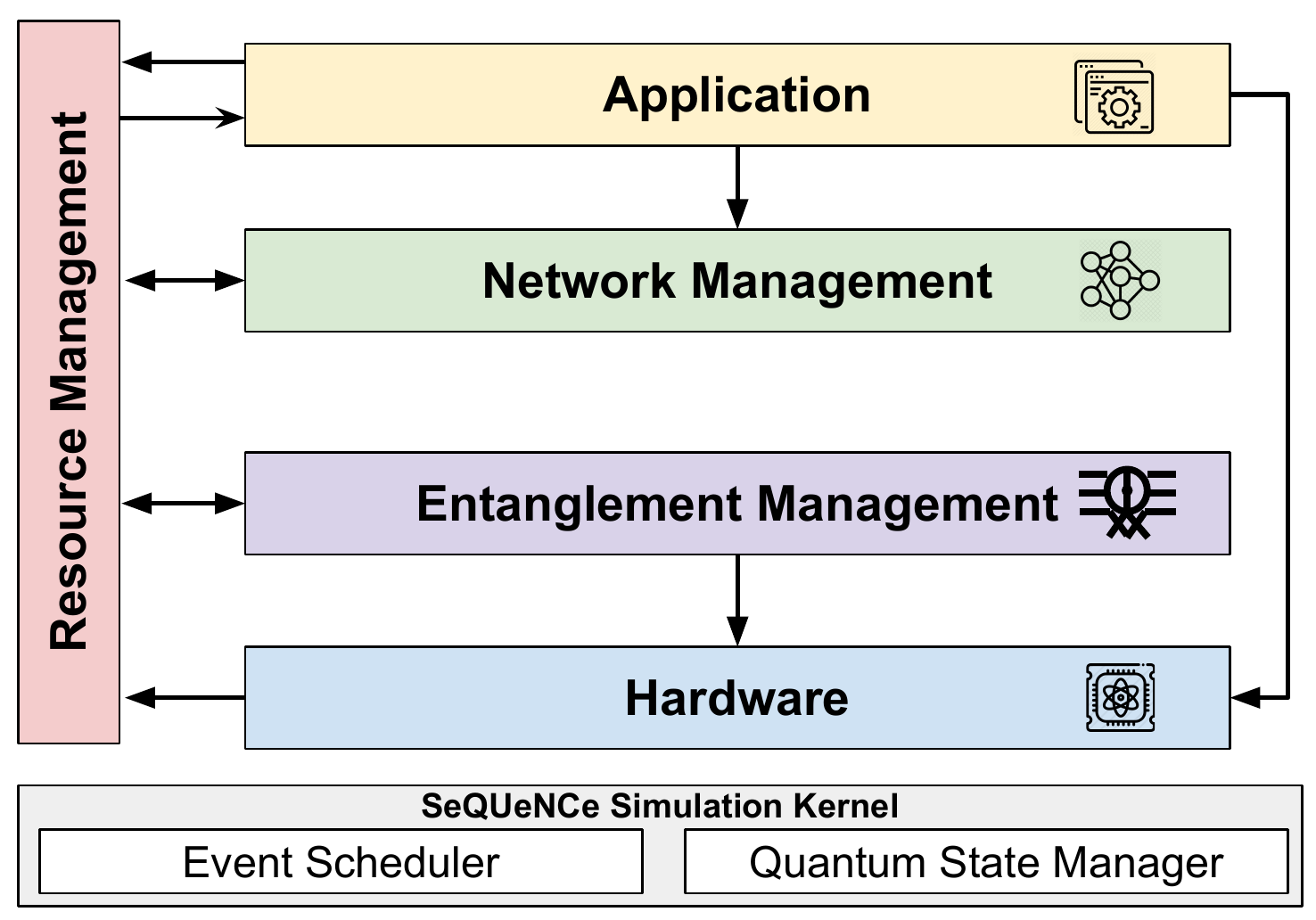}
  \caption{Modularized design of SeQUeNCe.}
  \label{fig:seq_arch}
\end{figure}

SeQUeNCe is an open-source quantum network simulator~\cite{sequence} that allows simulations with photon-level details. The SeQUeNCe project website~\cite{sequence-github} provides the complete source code, documentation and tutorials that allow easy usability and extendability. Among the many use cases of the simulator are understanding of the tradeoffs of alternative quantum network architectures, optimizing quantum hardware, and developing a robust control plane. SeQUeNCe uses a modularized design, as shown in Figure~\ref{fig:seq_arch}. This enables researchers to flexibly customize the simulated network and reuse existing models. The Application module represents quantum network applications and their service requests. The Resource Management and Network Management modules use classical control messages to manage allocation of network resources. The Entanglement Management and Hardware modules encompass protocols and hardware operations that act on qubits that are located in quantum memories or in photons at telecommunication wavelengths in optical network links. The SeQUeNCe simulator stores these states in the Quantum State Manager (QSM). For sequential simulation, a local QSM is sufficient to trace the quantum states. However, parallel simulation requires the simulator to synchronize multiple QSMs, a requirement that is addressed in this work. This work also significantly redesigned the Simulation Kernel to allow parallel simulations. Further details about the design and use cases of the sequential version of the SeQUeNCe simulator can be found in our earlier work~\cite{sequence, wu2019simulations, spw_19}.

\subsection{Parallel Simulation of Classical Networks} \label{sec:classical-netsim}
Researchers have built time-parallel and space-parallel network simulators to improve scalability in terms of both execution speed and network size. Time-parallel simulation methods~\cite{andradottir95,Jones01,wu03} parallelize a fixed-sized problem by partitioning the simulation time axis into intervals and assigning a processor to simulate the system over its assigned time interval.
Parallel (and distributed) discrete-event methods use a space-parallel approach that divides a single simulation topology into distinct pieces and executes them in parallel on multiple interconnected processors. These efforts started about 15 years ago with Parallel/Distributed ns~\cite{pdns} and the Georgia Tech Network Simulator~\cite{gtnets}. Distributed simulation support was added to ns-3~\cite{ns3} in 2011~\cite{Pelkey11}.

\begin{figure}
    \centering
    \includegraphics[width=0.5\textwidth]{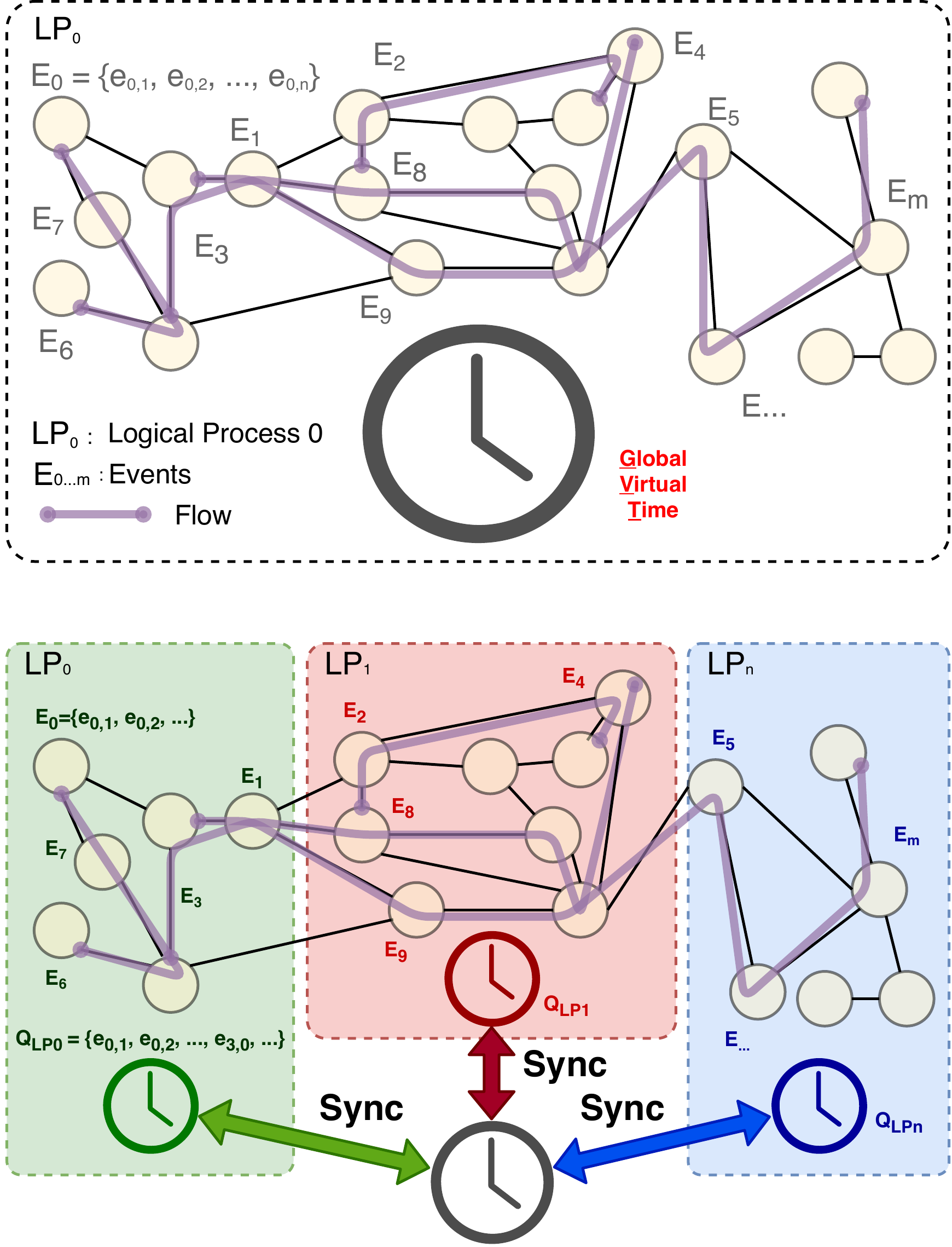}
    \caption{The causality constraint in parallel and distributed discrete-event simulations ensures that events are processed in time-stamped order. The insert on the top shows the sequential simulation of a classical network with a global virtual time, while the bottom part shows the same network topology simulated in parallel. The topology and events are partitioned and assigned to three processes, each with its corresponding local virtual time that has to be synchronized with the global virtual time.}
    \label{fig:netsim-into}
\end{figure}

Ensuring that events are processed in time-stamped order is a fundamental part of distributed simulation---the so-called causality constraint; see Figure~\ref{fig:netsim-into}. Almost all parallel network simulators use a conservative synchronization algorithm that takes precautions to avoid violating the causality constraint. The synchronization overhead significantly limits the speedup that can be achieved with parallel and distributed simulations. In fact, synchronization overheads in conservative approaches can cause parallel execution to be slower than sequential execution~\cite{kazer2018fast}. In contrast, optimistic synchronization algorithms allow violations to occur but are able to detect and recover from them.  Optimistic synchronization has been applied successfully in other contexts such as high-performance computing (HPC), storage, and HPC interconnects~\cite{carothers2002ross,codes} but has not been used to simulate wide area networks and TCP/IP-based networks (to the best of our knowledge). 

%% file: design.tex
\section{Parallel Quantum Network Simulator Design}
\label{sec:design}

The parallel simulation splits the simulated quantum network into multiple non-overlapping subnetworks. Each subnetwork consists of a subset of the network nodes, including components of the node and associated qubits, that are executed by one processor. The design of the parallel quantum network simulator should satisfy three requirements.

\begin{enumerate}
    \item \textbf{Accuracy}: The  results of the parallel simulation should faithfully reproduce results from the sequential simulation. Parallelization should not sacrifice realism.
    \item \textbf{Scalability}: The parallelized simulator should have good scalability. The design must carefully address the parallelization overhead, otherwise the parallel simulator performance could be suboptimal and in the extreme case even worse than that of its sequential counterpart.
    \item \textbf{Usability}: The parallel simulator should be compatible with models used for sequential simulation. This requirement aims to reduce errors as well as allow switching between parallel and sequential simulation, thereby increasing the usability and flexibility of the simulator.

\end{enumerate}

In this work we extend SeQUeNCe~\cite{sequence} to a parallel simulator and demonstrate how the design achieves all three requirements.

\subsection{Requirement Analysis}
\label{subsec:requirements}

The \textbf{accuracy} of the simulation depends on the correct order of event execution and the accurate modeling of network state. During a parallel simulation, a large network is partitioned into multiple non-overlapping subnetworks. Events in a subnetwork are executed by one assigned processor. Event execution may affect the order of events and the state of the network on the other processors. The accuracy of the simulation requires the simulator to synchronize these effects among processors. We summarize three methods that can affect the state of nodes on the other subnetworks. 

\begin{enumerate}
    \item One node can send qubits to another node over a quantum channel. This action generates an event on the receiver node to receive qubits. This is typically used for entanglement generation between adjacent quantum routers.
    \item One node can send classical messages to another node on a classical channel. This action creates an event to receive messages. Control protocols on nodes process these classical messages. 
    \item Entangled qubits that belong to different processors can change the shared entanglement state. For example, in quantum teleportation, the measurement on the sender node changes the quantum state of the qubit on the receiver node without any explicit quantum or classical message transmission.
\end{enumerate}

The first two methods affect states on other processors by generating events in the future. Processors can synchronize their states by exchanging cross-process events, and the synchronization algorithm can successfully avoid out-of-order event execution. The third method affects states on other processors by changing a shared state. To ensure consistency of shared states, we make entangled qubits share a single entanglement state instance instead of storing multiple copies for every qubit. A read/write lock is applied to avoid conflict when accessing those shared states.

Random number generation is another factor affecting reproducibility. Given the randomness in measuring qubits, the parallel simulator should generate random numbers identical to those of the sequential simulation. Therefore, we assign a random number generator (RNG) to every network node both in the sequential and parallel simulation. Events executed on a node or its components use the RNG on the node to generate a random number. Since components connected to a single network node are never split between processes, identical results will be produced as long as events on the node are executed in the correct sequential order.

The \textbf{scalability} of the simulator depends on the reduction of parallelization overhead. In a parallel quantum network simulator, the overhead results from synchronizing processors, exchanging data, and accessing the shared quantum states. Our design aims to reduce the execution time and complexity of these functions in order to improve the simulation scalability.

The \textbf{usability} of the parallel simulator depends on its ease of use. Models developed for parallel simulation ought to be compatible with the sequential simulation (one can configure the parallel simulator to run experiments with one processor). Therefore, our design should allow efficient and error-free simulation design; in other words, the parallelization should be systematically handled by the simulation kernel, with the process completely transparent to users.

\subsection{Parallel Simulator Architecture}

\begin{figure}[h]
  \centering
  \includegraphics[width=0.5\linewidth]{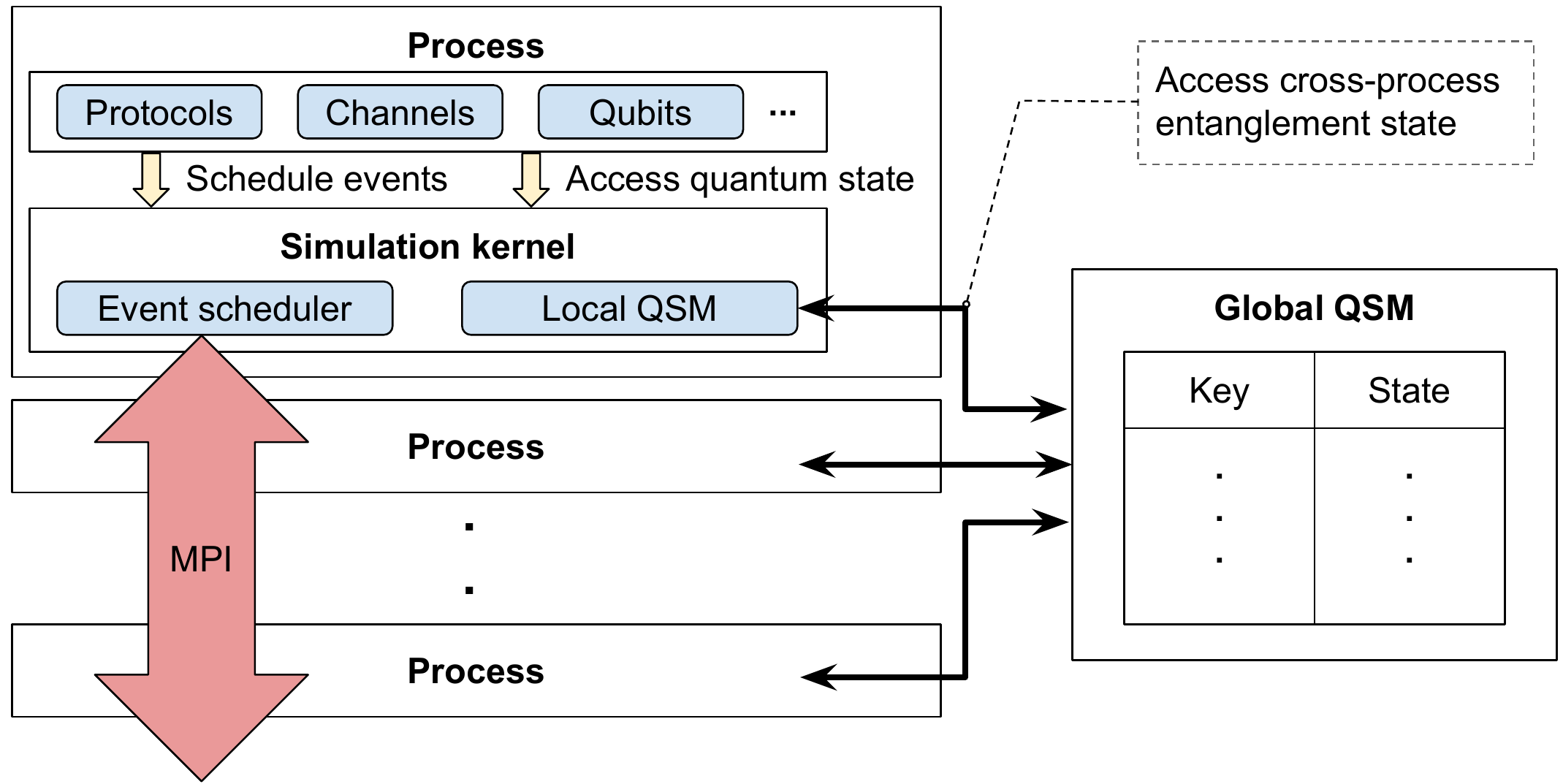}
  \caption{Architecture of the parallel simulator. Processes communicate and synchronize using MPI. Inside each process, models use the simulation kernel to schedule events and access quantum states. Local QSMs in each kernel forward cross-processes distributed entanglement states to the global QSM.}
  \Description{Two-layer structure where the first layer is the quantum state manager server and the second layer contains MPI processes.}
  \label{fig:arch}
\end{figure}

Figure~\ref{fig:arch} depicts the architecture of our parallel simulator. We parallelize SeQUeNCe by multiprocessing, where a separate processes handles each subnetwork. We use MPI~\cite{mpi} for interprocess communication  including cross-process events, produced by cross-process quantum and classical channels, for both shared- and distributed-memory systems. These two techniques allow users to run a  parallel simulation on an MPI cluster. Although  multiprocessing may incur a higher communication cost than multithreading does, high-performance computing cluster optimizations have reduced the MPI communication overhead~\cite{hpc_MPI}. Our experiments in Section~\ref{sec:eval} also demonstrate that MPI communication has a limited impact on the parallelization overhead.

To track simulated quantum states, we design a quantum state manager, which provides an interface for manipulating and reading quantum states. For shared entanglement, we also design a global QSM as a separate server. During the simulation, all cross-process entanglement states are stored in the global QSM, while non-entangled qubits or those entangled with only local qubits are stored in a local QSM. Entangled qubits  hold the key for accessing their shared state only within a local or global QSM. An individual process may then communicate with the global QSM to retrieve or update the state of an entangled qubit, either directly or through a quantum circuit. Quantum circuit operations, represented as unitary matrices, are not generally commutative. These operations therefore cannot be reordered, implying a requirement for proper ordering with respect to the simulation time for global QSM requests. Since quantum circuits may operate only on qubits stored in local hardware, however, circuits on separate processes will not overlap. Thus, the execution order of requests does not affect final results. The global QSM may be implemented as a single-process program, a multithreaded program, or a cluster according to the scale of the parallel simulation. Thus, the global QSM can handle requests from different processes more efficiently and reduce potential resource competition. While interprocess events are exchanged using MPI, the simpler point-to-point communication between the local QSMs and the global QSM utilizes TCP/IP communication. This is to avoid the additional overhead from using MPI while maintaining QSM communication. Further QSM implementation details are described in Section~\ref{sec:implementation}.

\begin{figure}[b]
\centering
    \begin{subfigure}[t]{0.47\linewidth}
        \centering
        \includegraphics[width=0.9\linewidth]{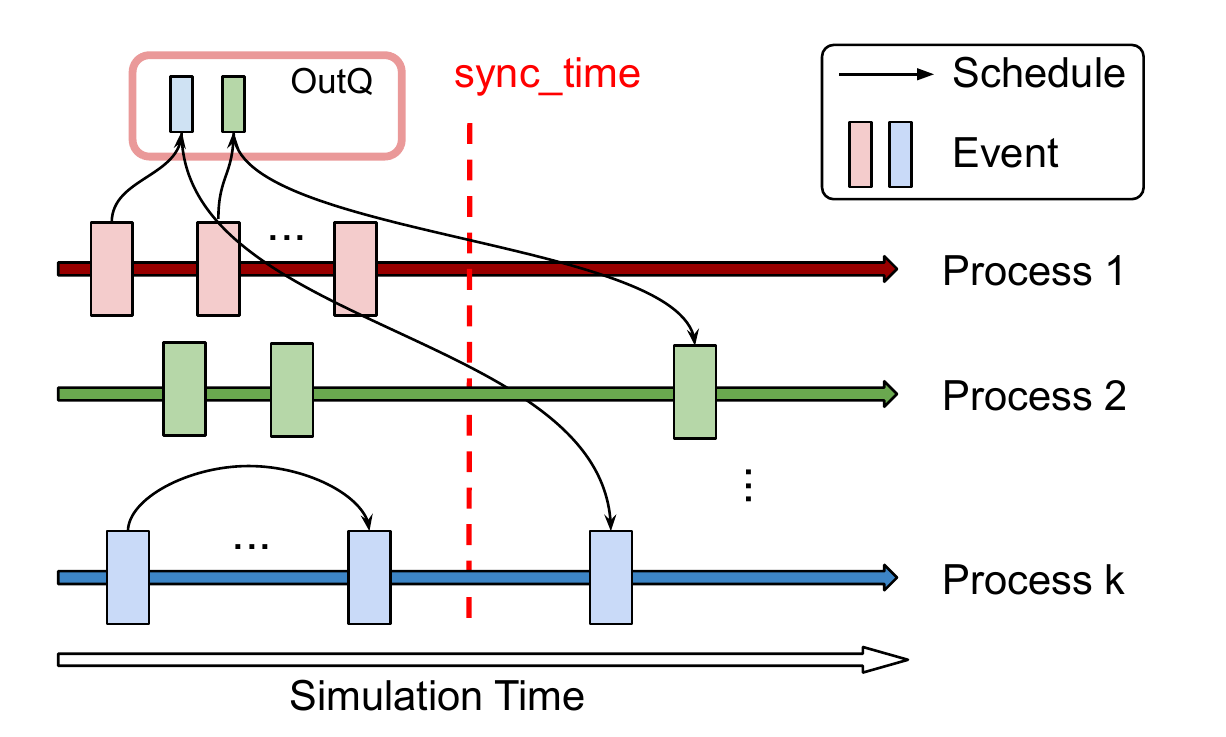}
        \caption{Parallel simulation from the perspective of simulation time. Processes independently execute events with a timestamp lower than the $sync\_time$. If the process generates events belonging to another process, these events are stored in the $OutQ$ of the process until all processes reach the $sync\_time$. The synchronization algorithm ensures that all events in each $OutQ$ have a timestamp that is larger than the $sync\_time$.}
        \label{fig:parallel_in_sim}
    \end{subfigure}%
\hspace{1em}
    \begin{subfigure}[t]{0.47\linewidth}
        \centering
        \includegraphics[width=0.8\linewidth]{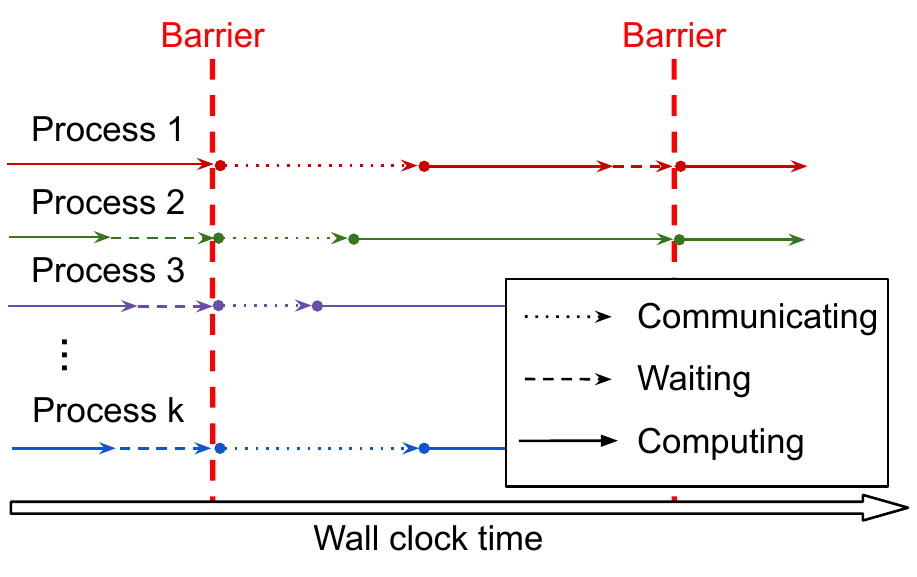}
        \caption{Parallel simulation from the perspective of wall-clock time. During the $Computing$ phase, processes independently execute events until the barrier required for the collective communication using MPI. During the $Communicating$ phase, processes exchange events and decide the next $sync\_time$ using MPI communication. During the $Waiting$ phase, processes wait for the slowest process to reach the synchronization barrier.}
        \label{fig:parallel_in_real}
    \end{subfigure}
    \caption{Conservative synchronization algorithm.}
  \centering
  \label{fig:conserv_exec}
\end{figure}

To ensure proper execution of events, we adjust the conservative synchronization algorithm~\cite{conservative} not only to exchange cross-process events but also to synchronize states on the global QSM. The existing conservative algorithm has been demonstrated to work for networks without entangled states. Figure~\ref{fig:conserv_exec} depicts the operation of the conservative algorithm. Events on separate processes execute in parallel during the $Computing$ phase. If an event execution schedules a future event to be run on a separate process, the future event is stored in the output queue $OutQ$. When a process executes all local events up to  $sync\_time$, it enters the $Waiting$ phase until all processes are finished. The processes then exchange events in the $OutQ$ during the $Communicating$ phase. Additionally, a future $sync\_time$ is calculated by using a lookahead value.

We describe our implementation of the conservative algorithm with the addition of a global QSM in Algorithm~\ref{alg:conservative}. In this algorithm, we introduce the following data structures:
\begin{itemize}
    \item $Queue$: a queue storing events to be executed locally
    \item $OutQ$: a queue of events that are scheduled locally but should be executed on a different process
    \item $InQ$: a queue of events that are scheduled by a different process but should be executed locally
\end{itemize}

\begin{algorithm}[]
        \SetAlgoLined
        
        \SetKwFunction{min}{min}\SetKwFunction{mpiex}{MPI\_EXCHANGE\_DATA}\SetKwFunction{push}{push\_events\_to\_queue}\SetKwFunction{exec}{execute}\SetKwFunction{notify}{notify\_server\_about\_synchronization}
        
		\While{local\_time < end\_time}{
		    \tcp{Communicating phase}
		    $local\_min\_time \leftarrow$ \min{$\{Queue.min\_time(), OutQ.min\_time()\}$}
		    
            $InQ, remote\_min\_times \leftarrow$ \mpiex{$OutQ, Local\_min\_time$}
            
            $global\_min\_time \leftarrow$ \min{$\{Local\_min\_time\} \cup remote\_min\_times$}
            
            $sync\_time \leftarrow$ \min{$\{global\_min\_time + lookahead, end\_time\}$}
            
            \push{$Queue, InQ$}
            
            \If{global\_min\_time is $\infty$}{
                break
            }
            
            \tcp{Computing phase}
            \While{$Queue.size() > 0$ \textup{\textbf{and}} $Queue.top().time < sync\_time$}{
                $event \leftarrow Queue.pop()$
                
                $local\_time \leftarrow event.time$
                
                \exec{$event$}
            }
            
            $local\_time \leftarrow sync\_time$
            
            \notify{}
		}
    	\caption{Conservative Synchronization}
		\label{alg:conservative}
\end{algorithm}

At each iteration of the algorithm loop, we check to see whether $local\_time$ (the simulation time of event execution on the process) is less than $end\_time$ (the simulation time at which event execution should halt). If the condition is met, the algorithm moves into the $Communicating$ phase; otherwise, the loop exits because the simulation has finished.

In the $Communicating$ phase, MPI processes first calculate the minimum timestamp of events stored locally ($Local\_min\_time$) by comparing the minimum timestamp of the events in the queue ($Queue$) and output queue ($OutQ$). Next, MPI processes move events in the local output queue to the corresponding process and broadcast their $local\_min\_time$. The function ``MPI\_EXCHANGE\_DATA'' returns events from other processes to the input queue ($InQ$) and a list of $local\_min\_time$ from other processes in the $remote\_min\_times$ list. By calculating the minimum value of the  $remote\_min\_times$ list and  $local\_min\_time$, the algorithm determines  $global\_min\_time$, which is the minimum timestamp of all events in the simulation. If all processes have an empty event queue, $global\_min\_time$ will be infinity, and execution is halted. Otherwise, we can use $global\_min\_time$ and  the $lookahead$ value to get the time window~\cite{conservative} bounded by a future $sync\_time$ (in simulation time). If the time window extends beyond $end\_time$, we instead set $sync\_time$ to be $end\_time$.

As shown in Figure~\ref{fig:parallel_in_sim}, the $lookahead$ value is set so that all cross-process events may  be scheduled for simulations times only after $sync\_time$. We subsequently move into the $Computing$ phase, where events in the local $Queue$ are executed and $local\_time$ is updated accordingly. After all events before $sync\_time$ are executed, we update $local\_time$ to be $sync\_time$. Additionally, the process sends a message to the global QSM to serve all pending requests. The process is unblocked upon  receiving a response from the global QSM to ensure that all the early requests on the global QSM are completely handled.

The parallel simulation kernel also supports sequential models for compatibility when needed. As shown in Figure~\ref{fig:arch}, models can schedule events and manipulate quantum states using interfaces offered by the kernel. When the simulation is executed sequentially, the event scheduler does not need to communicate with other processes, and the local QSM does not need to connect with the global QSM. Therefore, the same model can be executed using sequential or parallel simulation with only a simple setup configuration change.

\subsection{Parallelization Overhead Sources}
\label{subsec:overhead}

The proposed architecture introduces parallelization overhead. This overhead may be measured as an increase in execution time beyond that required for execution of simulated events. Figure~\ref{fig:parallel_in_real} shows the state of processes using the conservative synchronization algorithm. Processes have three states: $Computing$, $Communicating$, and $Waiting$. The $Computing$ state denotes that the process is executing events whose timestamps are within the time window. The $Communicating$ state denotes the MPI communication that creates a synchronization barrier among processes. The $Waiting$ state denotes that a process has completed the local event execution and is now blocked by the communication barrier. We identify four sources of overhead. First, the barrier caused by the MPI communication makes a process (e.g., light workload) wait for other processes (e.g., heavy workloads) in the $Waiting$ state. The second source is due to the data exchange among processes. The number of exchanged events affects the communication overhead. The third source is the cost for barrier synchronization (e.g., the calculation of the local minimum timestamp for the next computing phase and the time window until the next synchronization). Apart from the overhead for the synchronization algorithm, the global QSM introduces the fourth overhead source. In the parallel simulation, the global QSM handles operations on a cross-process entanglement state. As a result, the communication spent on the global QSM increases the time of manipulating entanglement states. A parallel simulation with a larger amount of cross-process entanglement may thus introduce larger parallelization overhead due to the growing global QSM communication. In Section~\ref{sec:implementation} we introduce multiple techniques for reducing the overhead from these sources.

%% file: implementation.tex
\section{Quantum State Manager}
\label{sec:implementation}

This section describes the quantum state manager for our PDES-based simulator of quantum networks. Section~\ref{subsec:state_manager} describes the essential functions provided by the QSM in the simulation kernel and the memoization technique  used for accelerating matrix multiplication. Section~\ref{subsec:server} describes the global QSM, in which we explain the mechanism used for distinguishing cross-process entanglement and single-process entanglement. Additionally, we illustrate the local computing, traffic compression, and multithreading techniques that are used to reduce the overhead of communication between the local and global QSMs.

\subsection{Local Quantum State Manager}
\label{subsec:state_manager}

The QSM within the simulation kernel is designed to provide methods for manipulating quantum states. Models of hardware and protocols may manipulate the quantum state of qubits with a qubit key that is produced by a universal unique identifier~\cite{uuid}. The QSM maintains the relation between keys and quantum states as a map data structure. For entangled states, multiple keys may point to the same quantum state object. Here we introduce three important functions provided by the QSM:

\begin{itemize}
    \item $set(keys, state)$: The QSM sets the quantum state to a given state for qubit(s) with a given key(s).
    \item $get(key)$: The QSM returns the state of the qubit with a given key.
    \item $run(circuit, keys, prob\_sample)$: The QSM runs a quantum circuit on the qubit(s) with a given key(s). The circuit measures qubits, $prob\_sample$, a random number in the range from 0 to 1 for deciding the collapsed quantum state. The function returns the outcome of the measurement.
\end{itemize}

Quantum states (including entanglement) are stored as vectors (Bra-ket notation~\cite{ketstate}) or matrices (density matrix~\cite{density_matrix}) together with a list that indicates the order of entangled qubits as a list of keys. The $get$ and $set$ functions allow models to read and write the quantum state of qubits. Compared with the $set$ function, hardware and protocol models mostly use the $run$ function to change the quantum state of qubits. For example, entanglement purification~\cite{entanglement_purification} and swapping~\cite{swapping} protocols usually describe their protocols as a quantum circuit. A quantum circuit consists of quantum logic gates and/or measurement. The gates in the circuit may be represented by a unitary matrix. 
Unlike quantum logic gates, measurement is a stochastic nonreversible operation. The simulator thus needs to use a random number to determine the measured state. To reproduce identical outcomes, the $run$ function takes the random number $prob\_sample$ from the caller function if the circuit contains a measurement operation. As described in Section~\ref{subsec:requirements}, this random $prob\_sample$ is produced by the random number generator on the calling node to ensure consistency of simulations.

In SeQUeNCe, we use the QuTiP~\cite{qutip} Python library to generate the unitary matrix of a given circuit. The output state is produced by matrix multiplication of the unitary matrix of the circuit gates with the input quantum state of the qubit(s), which is a CPU-intensive task. Furthermore, the simulation of a quantum network involves repeated computation of this matrix multiplication. Therefore, we utilize the memoization technique~\cite{memoization} to speed up redundant functions, such as the $run$ function. We implement least recently used caches with fixed sizes to cache the result of these multiplications.

\subsection{Global Quantum State Manager}
\label{subsec:server}

While unentangled states or entangled states confined to one process may be managed by a single local QSM, cross-process entangled states may need to be accessed by multiple local QSMs. We thus designed a global QSM as a server to ensure the consistency and accuracy of quantum states within the parallel simulation. While the quantum state of qubits is stored in the centralized global QSM, the local QSMs on separate MPI processes can manipulate a shared state without conflict by using identical functions.

Since the global QSM is executed on an independent process, it must receive requests and send responses over network communication in contrast to a local QSM. Multiple MPI processes may communicate with the global QSM simultaneously, which can result in resource competition. Such competition may increase the processing time on the global QSM when performing large-scale parallel simulations. The communication time and resource competition constitute the parallelization overhead from the global QSM as described in Section~\ref{subsec:overhead}. The global QSM may thus become a bottleneck for parallel simulation. To address this, we implement multiple techniques to efficiently reduce the global QSM overhead.

\subsubsection{Mitigating Communication Overhead}

\begin{figure}[!]
    \centering
    \begin{subfigure}[t]{0.385\linewidth}
        \centering
        \includegraphics[width=\linewidth]{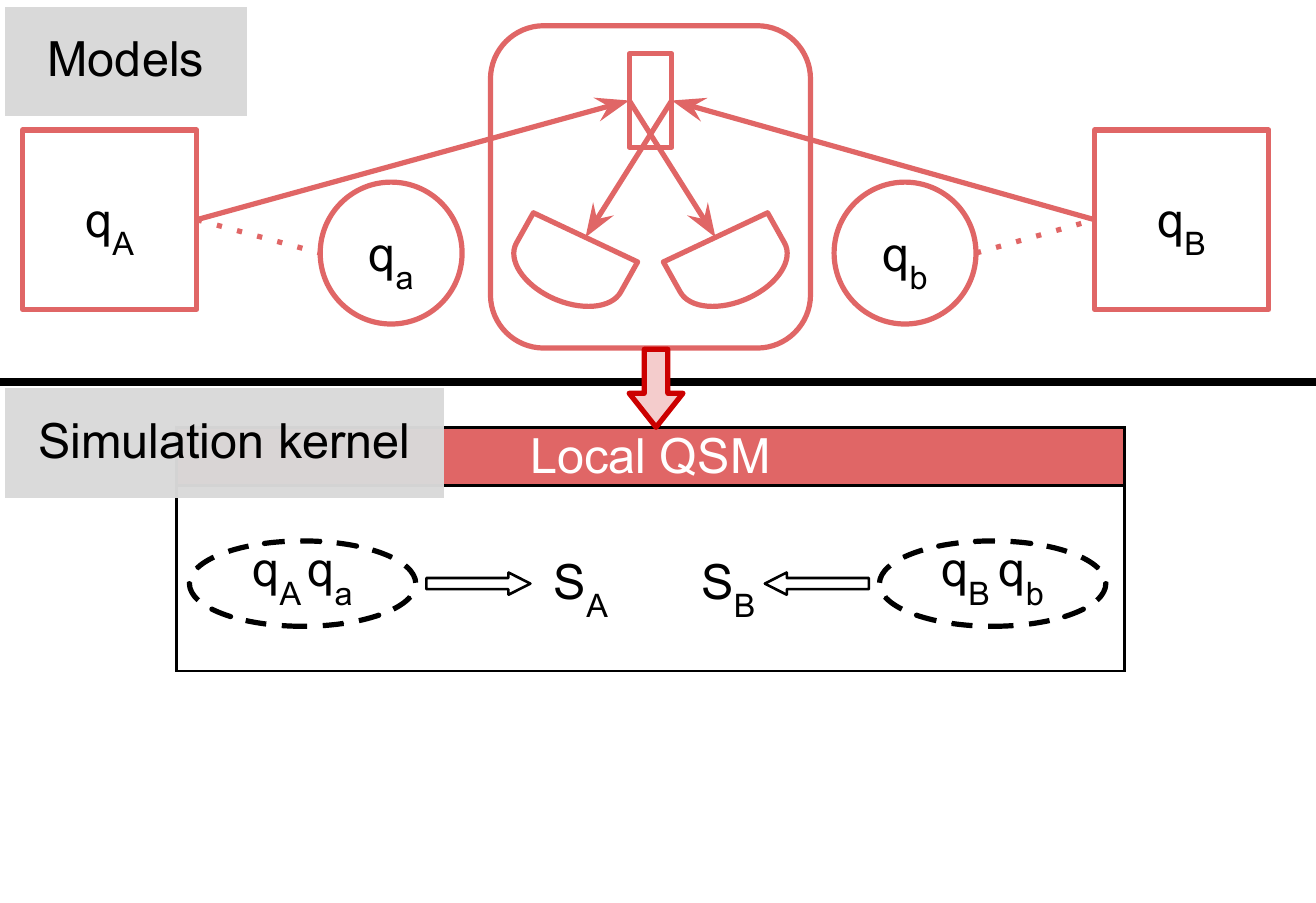}
        \caption{All qubits belong to one MPI process. The local QSM of the process updates the quantum state of qubits locally.}
        \label{fig:state_management_local}
    \end{subfigure}
    \hfill
    \begin{subfigure}[t]{0.55\linewidth}
        \centering
        \includegraphics[width=\linewidth]{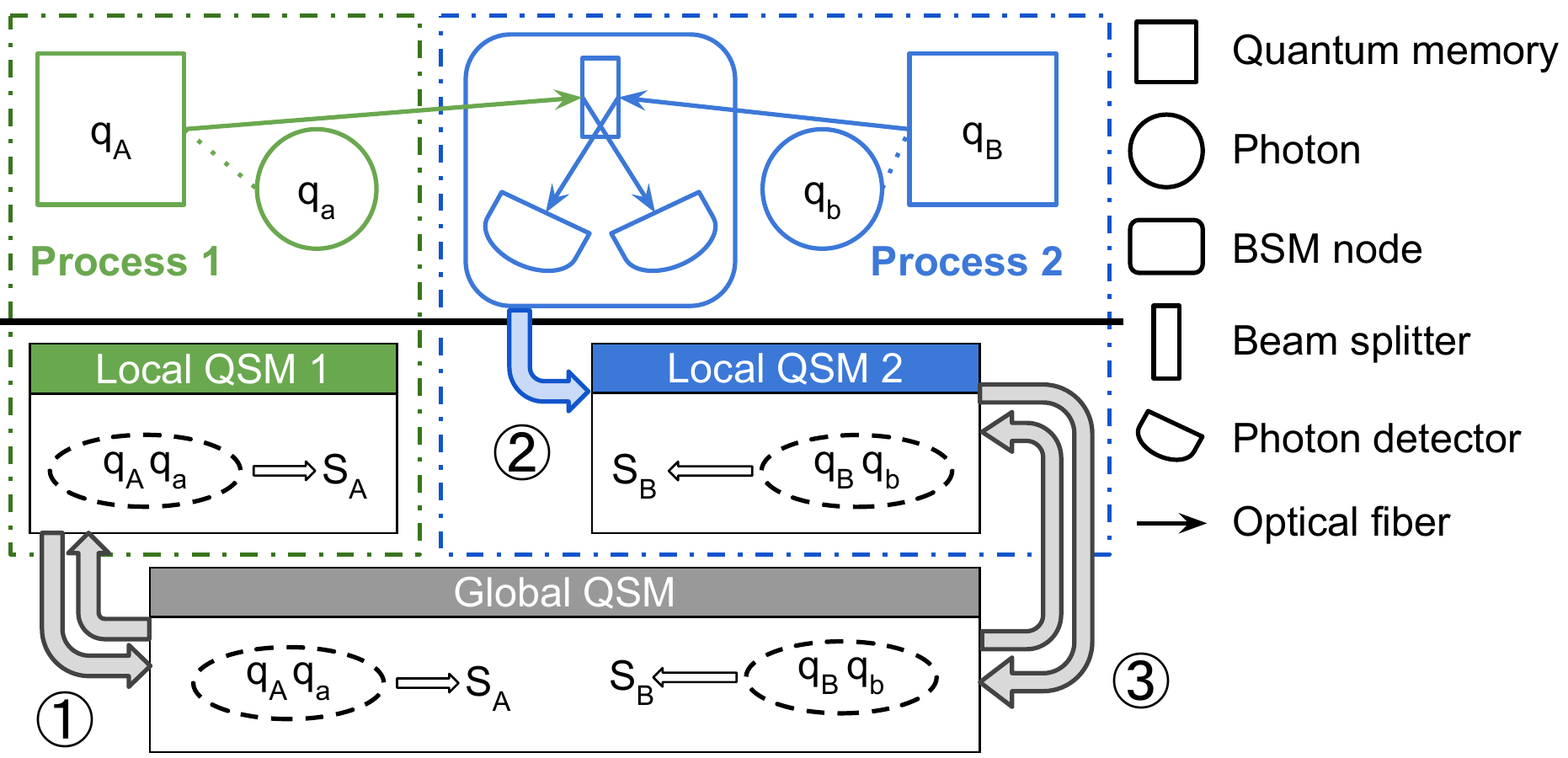}
        \caption{Qubits belong to different MPI processes; entities in the left (right) part belong to  process 1 (process 2); the photon $q_a$ moves from  process 1 to process 2. When process 1 detects that the destination of $q_a$ is on process 2, the local state is forwarded to the global QSM in \circled{1}. The optical hardware models on process 2 use the local QSM to generate entanglement in \circled{2}. The local QSM detects that a required qubit is stored on a separate process and forwards to the global QSM in \circled{3}. The global QSM then produces the required entanglement state and returns results to the local QSM on process 2.}
        \label{fig:state_management_remote}
    \end{subfigure}
    \caption{Entanglement generation with the meet-in-the-middle protocol under two scenarios; quantum memories $q_A$ and $q_B$ emit photons $q_a$ and $q_b$, respectively, to generate entanglement. $S_A$ ($S_B$) denotes the quantum state of qubits $q_A$ ($q_B$) and $q_a$ ($q_b$). The optical fiber transmits photons to the BSM node located in the middle. Once detectors are triggered by photons, the BSM node sends messages to nodes containing memories. Protocols may determine the current quantum state of memories with these messages.}
    \label{fig:state_management}
\end{figure}

The total communication time depends on the number of transmissions and the latency of each transmission.
In this work we reduced the number of transmissions between the local and global QSMs. These transmissions include three types of communication categorized by their goals:

\begin{itemize}
    \item \textbf{Forward function call}. The local QSM forwards function calls to the global QSM and waits for a response. 
    \item \textbf{Synchronize quantum state}. Operations on the local QSM may cause a state inconsistency, which requires the local QSM to update the state in the global QSM.
    \item \textbf{Acknowledge state of global QSM}. This is for a local QSM to confirm that the global QSM has completed  executing all the early forwarded functions. 
\end{itemize}

To reduce the number of transmissions, we first offload most of the computational work to the local QSM. We use the entanglement generation meet-in-the-middle protocol~\cite{generation_schemes} as an example to show interactions between the local and global QSM in Figure~\ref{fig:state_management}. Qubits $q_A$ and $q_B$ stored in each quantum memory are entangled with photons $q_a$ and $q_b$ with the shared quantum state $S_A$ ($S_B$). Photons are sent to the intermediate BSM node to generate entanglement between $q_A$ and $q_B$. In the parallel simulation, qubits either belong to one process, shown in Figure~\ref{fig:state_management_local}, or belong to two processes, shown in Figure~\ref{fig:state_management_remote}. If the MPI process contains all required qubits, the local QSM can produce the entangled state between $q_A$ and $q_B$ locally without forwarding the function to the global QSM. Otherwise, the local QSM must forward the function to the global QSM since the ``local QSM 2'' executing events on the BSM node does not hold the state $S_A$. As a result, a simulation with fewer cross-process entanglement states has fewer communication requests with the global QSM and thus reduces the communication overhead.

The challenge of offloading work to the local QSMs is to ensure the consistency of quantum states between the local and global QSMs. For example, in Figure~\ref{fig:state_management_remote}, the quantum state $S_A$ in the global QSM should be identical to the state $S_A$ stored in ``local QSM 1'' before the global QSM executes the request from ``local QSM 2.'' The timing of synchronizing local states with the global QSM thus becomes an issue, since the parallel simulator must keep this consistency while avoiding redundant synchronization.

Investigation of existing entanglement protocols shows that the synchronization communication is triggered by two conditions: 
(1) transmitted qubits sharing an entangled state and (2) forwarded function calls requiring a local quantum state. The first condition may separate two entangled qubits that belong to one process, so we synchronize the entanglement state to the global QSM. The management of this state is moved from the local QSM to the global QSM. To the best of our knowledge, all entanglement generation protocols rely on quantum channels to entangle qubits in two locations. The synchronization triggered by the first condition can avoid inconsistent states between the local and global QSMs. The second condition requires the global QSM to execute operations. The local QSM should push the local quantum state to the global QSM before forwarding the function call. The correctness of entanglement protocols, such as entanglement swapping, can be protected by these triggering conditions.

The synchronization requirement demands the management of qubits at the global QSM. The local QSM forwards the function calls related to either unknown qubits (i.e., qubits produced by other processes) or qubits managed by the global QSM.  Once the qubits lose their entangled state, the management of qubits will be transferred back to the local QSM. In the parallel SeQUeNCe simulator, either the measurement on qubits or the $set$ function can return the qubits back to the local QSM by destroying an entangled state.

In addition to offloading work from the global QSM, we  reduce the communication time by grouping multiple requests into one packet. We observe that some requests to the global QSM do not expect responses, such as the $run$ function without measurement and the $set$ function. Therefore, the simulator caches requests with no responses. These requests are transmitted to the global QSM only if a new request needs results from the global QSM or the simulator forcefully transmits all requests in the cache. This mechanism leads to efficient state synchronization since the synchronization does not require a response from the global QSM.

We implement the communication between the local and global QSM using the TCP/IP protocol. Functional calls are serialized in JSON format. We note that the communication latency highly depends on the performance of the MPI cluster. Thus, we execute the global QSM process on the same cluster to reduce the communication latency.

\subsubsection{Mitigating Resource Competition}

The resource competition of MPI processes reduces the performance of the global QSM. Since the global QSM needs to handle requests from all MPI processes, the performance of the global QSM will degrade with a growing number of processes and communication, and thus the global QSM may become the parallelization bottleneck.

To address this problem, we first design the global QSM as a multithreaded program. The global QSM uses independent threads to concurrently serve MPI processes. Compared with a single-process program, the multithreading technique reduces the waiting time  for executing the early arrival requests. Compared with a multiprocessing program, the multithreading technique provides a lighter way to share resources among threads.

Using the multithreading technique requires thread safety for the shared quantum states in the global QSM. In the parallel simulation of a quantum network, multiple MPI processes may simultaneously operate on the same quantum state. For example, step 2 of Figure~\ref{fig:es} shows that nodes A and C need to run a circuit to correct the quantum state of entanglement between A and C. If nodes A and C belong to two MPI processes, the two threads serving these MPI processes may simultaneously process requests that modify the entanglement state shared between A and C, which may cause a race condition. We use a set of locks to guarantee thread safety. The global QSM assigns each qubit a lock. When the global QSM receives a function call, the global QSM collects all qubit keys related to the function. These qubits include not only qubits listed in the argument of the function but also qubits entangled with listed qubits, since all these qubits share a single entangled state. Then, the thread acquires locks for these qubits in order to avoid deadlocks. After the execution of the function, the thread releases all locks.

In contrast to the rest of the SeQUeNCe source code, we implement the global QSM in C++. Since simulation models interact only with the local QSM, the global QSM is hidden from users and their models. Implementing the global QSM in C++ rather than Python results in higher performance while still maintaining the ease of use for the simulator. The C++ library Quantum++~\cite{quantum++} is used to manipulate quantum states on the global QSM. As an additional benefit, Quantum++ claims better performance than QuTiP~\cite{qutip}. The memoization mentioned in Section~\ref{subsec:state_manager} is also implemented to reduce redundant computational work.

%% file: lookahead.tex
\section{Lookahead Optimization}
\label{sec:async}

The lookahead value determines the frequency of synchronization among processes. Parallel simulations with larger lookahead values synchronize the processes less often and thus exhibit better performance. Additionally, the overhead from the unbalanced workload can be mitigated by a larger lookahead value. In order to support PDES of a quantum network, the minimum latency of cross-process channels can be selected as a baseline lookahead value. Although both quantum and classical channels introduce a propagation delay, the latency on the classical channels also includes processing delay, transmission delay, and queuing delay, which make the latency on a classical channel $T_{cc}$ generally larger than the latency on a quantum channel $T_{qc}$. Therefore, the lookahead value usually equals the latency of the shortest quantum channel.

However, we propose that the lookahead value $\Delta t$ could be increased to $\frac{T_{cc}}{2}$, where $T_{cc}$ is the classical delay between routers and intermediate BSM nodes if the simulated quantum network uses the meet-in-the-middle protocol to generate entanglement. Users could choose $\frac{T_{cc}}{2}$ as the lookahead value if $\frac{T_{cc}}{2} > T_{qc}$.  In a general entanglement distribution network, only the entanglement generation protocol relies on the quantum channel;  the remaining protocols manipulate entanglement using quantum circuit operations on a single node. When the longer latency is chosen, the simulator can  guarantee the accuracy of simulating a network only with the meet-in-the-middle configuration. Figure~\ref{fig:state_management} depicts the process of the meet-in-the-middle entanglement generation protocol. One BSM node is placed in the middle of two quantum routers. Routers first send qubit-encoded photons to the BSM node and then receive messages from the BSM node. In a parallel simulation, if two quantum routers belong to different processes, the BSM node is the entity that connects with two routers over the quantum channel. Routers using the meet-in-the-middle protocol send photons to the BSM node with $T_{qc}$ delay. The message events generated by the BSM node occur after a further delay of $T_{cc}$. We can use this transmission pattern to increase the lookahead value.

\begin{figure}
\centering
    \begin{subfigure}[t]{0.5\linewidth}
        \centering
        \includegraphics[width=\linewidth]{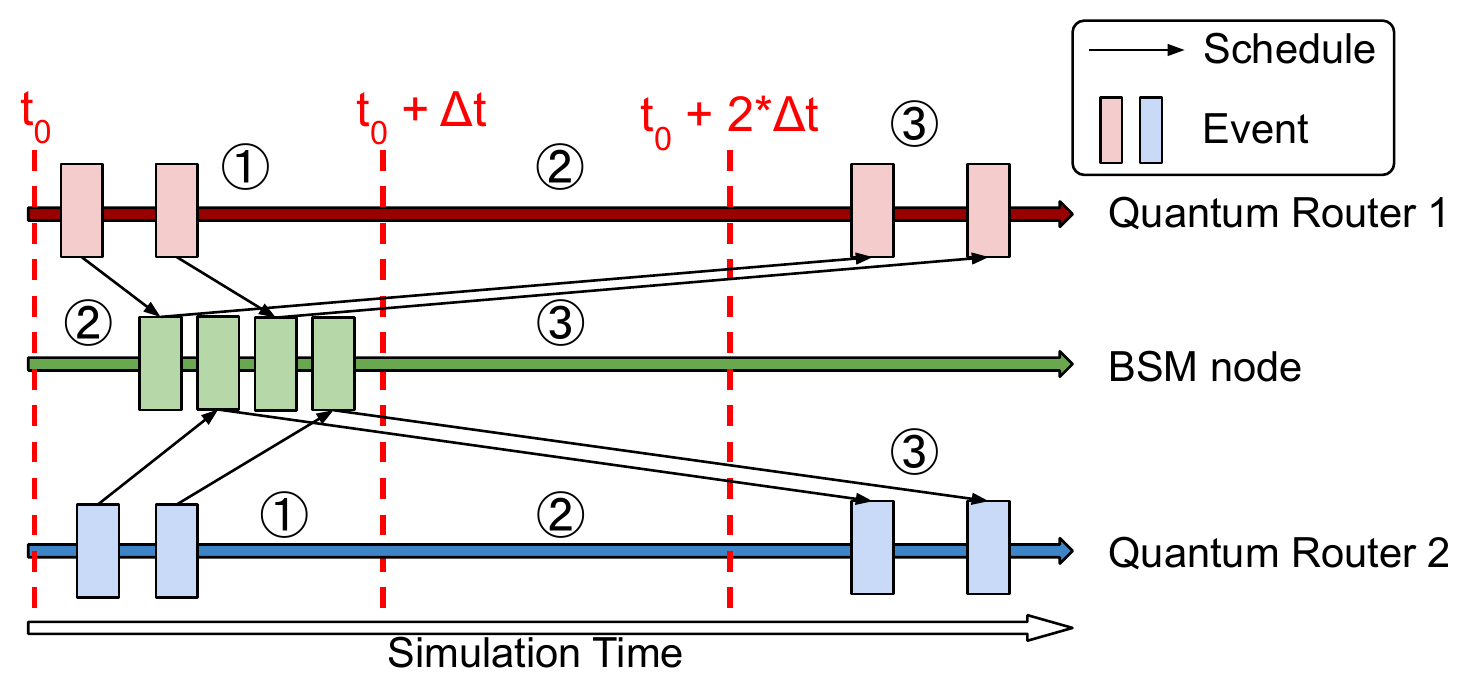}
    \caption{Event execution in the perspective of simulation clock time}
    \label{fig:async1}
    \end{subfigure}%
    ~ 
    \begin{subfigure}[t]{0.5\linewidth}
        \centering
        \includegraphics[width=0.8\linewidth]{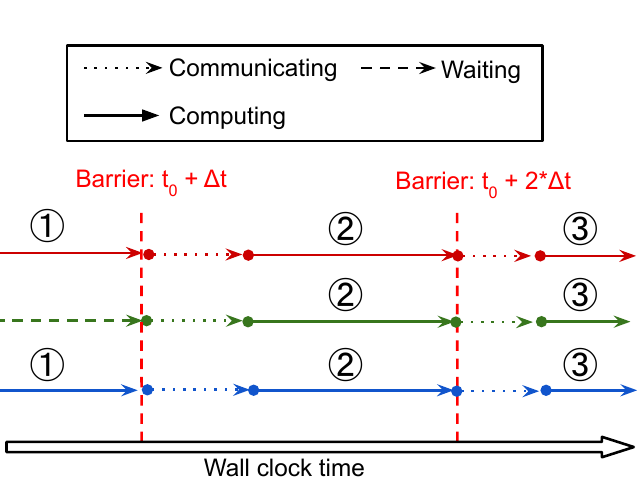}
        \caption{Event execution in the perspective of wall-clock time}
        \label{fig:async2}
    \end{subfigure}
    \caption{Parallel simulation using half of the classical channel latency as the lookahead value $\Delta t$}
  \centering
\end{figure}

Figures~\ref{fig:async1} and \ref{fig:async2} illustrate parallel simulation of the  meet-in-the-middle protocol with $\Delta T = \frac{T_{cc}}{2}$ and asynchronous processes. The simulation starts from  $t_0$, and three nodes maintain their time and events separately. Quantum routers synchronize their time by Algorithm~\ref{alg:conservative}, and the BSM node  exchanges cross-process events only at the time window. We label the order of execution on the figure as follows:
\begin{enumerate}[label=\protect\circled{\arabic*}]
\item Quantum routers execute events from $t_0$ to $t_0 +\Delta t$ and exchange events. Cross-process events are moved to the BSM node.
\item The BSM node executes events from $t_0$ to $t_0 +\Delta t$, while routers execute events from $t_0+\Delta t$ to $t_0 + 2 * \Delta t$. The BSM node sends produced cross-process events to the routers and receives cross-process events from $t_0 +\Delta t$ to $t_0 + 2*\Delta t$.
\item The BSM node executes events from $t_0 + \Delta t$ to $t_0 + 2*\Delta t$, and the routers execute events from $t_0 + 2*\Delta t$ $t_0 + 3 * \Delta t$. Then, the  nodes exchange data as described in \circled{2}.
\end{enumerate}
The BSM node receives events at the barrier $t_0 + \Delta t$ and executes events from simulation time $t_0$ to $t_0 + \Delta t$. Since the time of the BSM node is slower than the time on the routers by at most $2 \Delta t$, cross-process events produced by the BSM node must have a timestamp larger than  $t_0 + 2*\Delta t$. The second synchronization barrier at $t_0 + 2*\Delta t$ can move cross-process events produced by the BSM node to the routers. In this way, all cross-process events are exchanged and executed in the correct order, and PDES can use the $\Delta T = \frac{T_{cc}}{2}$ lookahead time to reduce the parallelization overhead if $\frac{T_{cc}}{2} > T_{qc}$. 

To apply this mechanism, the MPI process picks out BSM nodes that connect with routers on other processes before running the simulation. The MPI process maintains two simulation times and event queues for the selected BSM nodes and other nodes, respectively. The simulation times for the selected BSM nodes are allowed to be slower than for the  others. MPI processes then execute events as shown in Algorithm~\ref{alg:conservative} with the mechanism described previously to run the parallel simulation. We evaluate the improvement from this optimization in Section~\ref{sec:eval}.

%% file: evaluation.tex
\section{Performance Evaluation}
\label{sec:eval}
We conduct extensive evaluation experiments for the parallelized SeQUeNCe on three quantum network topologies. We report four components of the wall-clock execution time of the simulations:  MPI communication, socket communication, event execution, and barrier synchronization. The MPI communication includes the overhead from the data exchange and the unbalanced workload. The socket communication denotes the aggregated pending time of a process caused by the communication between the local and global QSMs. The execution time speedup (strong scaling) and efficiency (weak scaling)~\cite{scaling}, $T_s / T_p$, are used as the metrics to quantify the scalability of the simulator, where $T_s$ and $T_p$ denote the execution time used for sequential simulation and parallel simulation consisting of $p$ processes, respectively. The barrier synchronization algorithm overhead is very low (less than 1 second), and thus we do not plot it in the following figures.

The simulations of the three quantum networks utilize identical models of software and hardware. Simulations execute up to 30 million events by using 1-~128 processors. Considering the scale of the simulated network, we limit the maximum number of processors to 128, but our parallel simulator supports parallel simulations using more processors. In Table~\ref{tab:parameters} we list values of parameters shared among simulations;  the detailed model design can be found in our prior work~\cite{sequence}. For classical communication, the nodes are connected with the channels in a fully connected topology.

The three simulated quantum networks have different network topologies and flows. In Section~\ref{subsec:chain} the first quantum network uses a linear topology to distribute entanglement between the end nodes. Such a chain of quantum repeaters is useful for long-distance distribution of entanglement. The network uses only one traffic flow to distribute entanglement between the two routers on the endpoints. 

In Section~\ref{subsec:caveman} the second quantum network uses the topology of a connected caveman graph~\cite{caveman} that consists of 128 connected cliques, each having 8 quantum routers. This topology is more complicated than the linear topology. We simulate multiple random flows in the network. The random flows are produced by the network nodes, and each node selects a random destination.

The third quantum network uses the topology of an Internet autonomous system network~\cite{as_net} generated by Networkx~\cite{networkx}. Similar to the second network, every network node produces a flow with a random destination. The challenge of parallel simulation for this kind of network results from its irregular topology. We introduce three heuristic methods to optimize the network partitioning in Section~\ref{subsec:as_net}.

All the simulations were executed on the Broadwell partition of Bebop~\cite{bebop}, the high-performance computing cluster operated by the Laboratory Computing Resource Center at Argonne National Laboratory. Each of the Broadwell nodes has 36 cores of Intel Xeon E5-2695v4 and 128 GB of DDR4 running CentOs 7. All programs are interpreted using Python 3.8.5 or compiled using g++ 4.8.5. 

Bebop nodes use an Intel Omni-Path interconnect for communication, with a local MPI implementation. The other libraries used for simulation were NumPy (version 1.18.5), pandas (version 0.24.2), QuTiP (version 4.6.0.dev)~\cite{qutip}, tqdm (version 4.47.0), NetworkX (version 2.4), mpi4py (version 3.0.3), and Quantum++ (version 2.6)~\cite{quantum++}. Python packages were managed  using Anaconda version 4.8.3, and the C++ QSM server was built using CMake version 3.18.4.

\begin{table}[]
\caption{Parameters of hardware models and their default values.}
\begin{tabular}{|l|l|}
\hline
\textbf{Parameter}                                                          & \textbf{Value}      \\ \hline
Memory efficiency                      & $\eta_m = 0.75$          \\ \hline
Memory frequency                               & $f_m = 20$ KHz      \\ \hline
Memory coherence time                      & $t_c = 1.3$ s       \\ \hline
Fidelity of entanglement from generation protocol                           & $F_0 = 0.99$           \\ \hline
Detector efficiency                            & $\eta_d = 0.9$        \\ \hline
Detector count rate                      & $r = 25$ MHz        \\ \hline
Detector dark count             & $d  \approx 0~s^{-1}$          \\ \hline
Detector resolution                       & $s = 150$ ps        \\ \hline
Attenuation                    & $\alpha_0 = 0.2$ dB/km   \\ \hline
Speed of light                                  & $c^* = 2 \times 10^8$ m/s \\ \hline
Channel TDM time frame                       & $t_f = 12.5$ ns            \\ \hline
Gate fidelity                                                & $F_{gate} = 1$                \\ \hline
Swap success probability                 & $p_{swap} = 1$         \\ \hline
Quantum channel length  & $L_{qc}$ = 1 km \\ \hline
Classical channel latency & $T_{cc}$ = 0.3 ms \\ \hline
Simulation time & 100 ms \\ \hline
\end{tabular}
\label{tab:parameters}
\end{table}

\subsection{Chain of Quantum Routers}
\label{subsec:chain}

\begin{figure}[]
    \centering
    \begin{subfigure}[t]{0.4\linewidth}
        \centering
        \includegraphics[width=\linewidth]{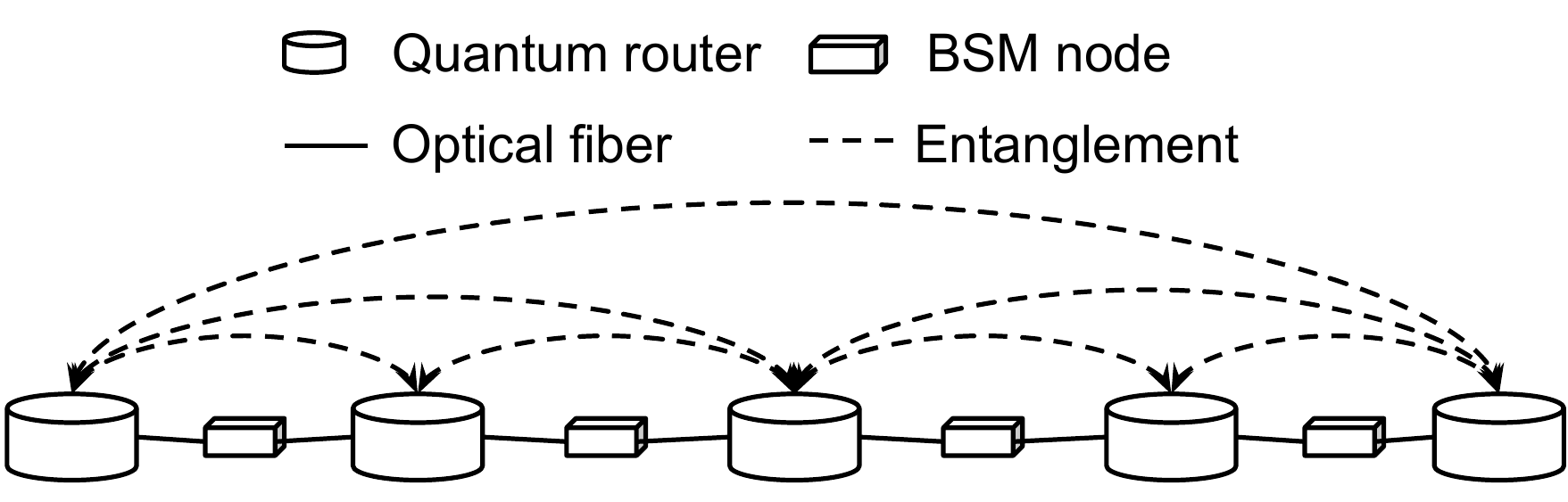}
        \caption{A small-scale linear network topology example with five routers}
        \label{fig:chain_topo}
    \end{subfigure}%
    \\
    \begin{subfigure}[t]{0.35\linewidth}
        \centering
        \includegraphics[width=\linewidth]{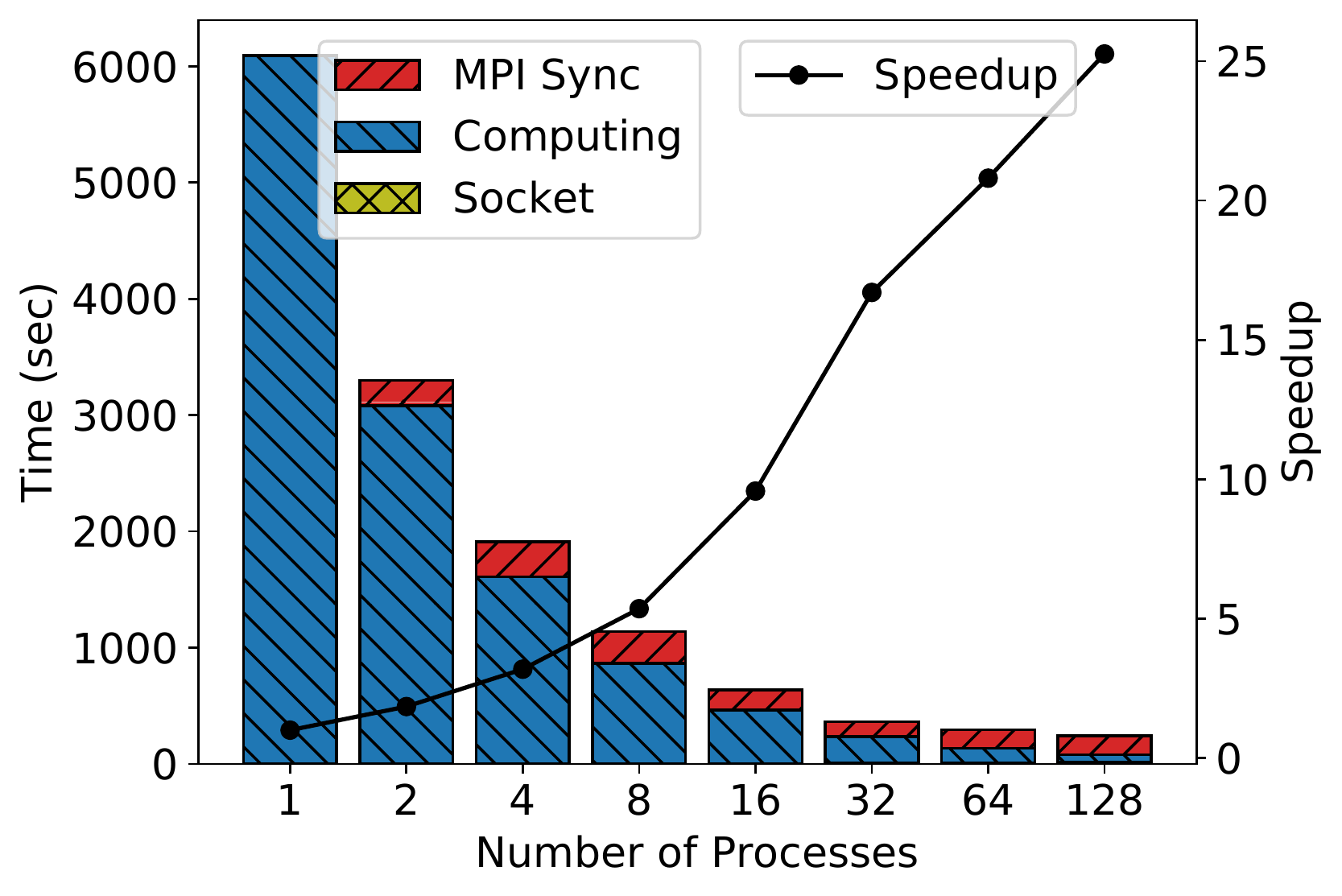}
        \caption{Execution time and speedup -- strong scaling}
        \label{fig:motivation_strong}
    \end{subfigure}%
    \hfill 
    \begin{subfigure}[t]{0.35\linewidth}
        \centering
        \includegraphics[width=\linewidth]{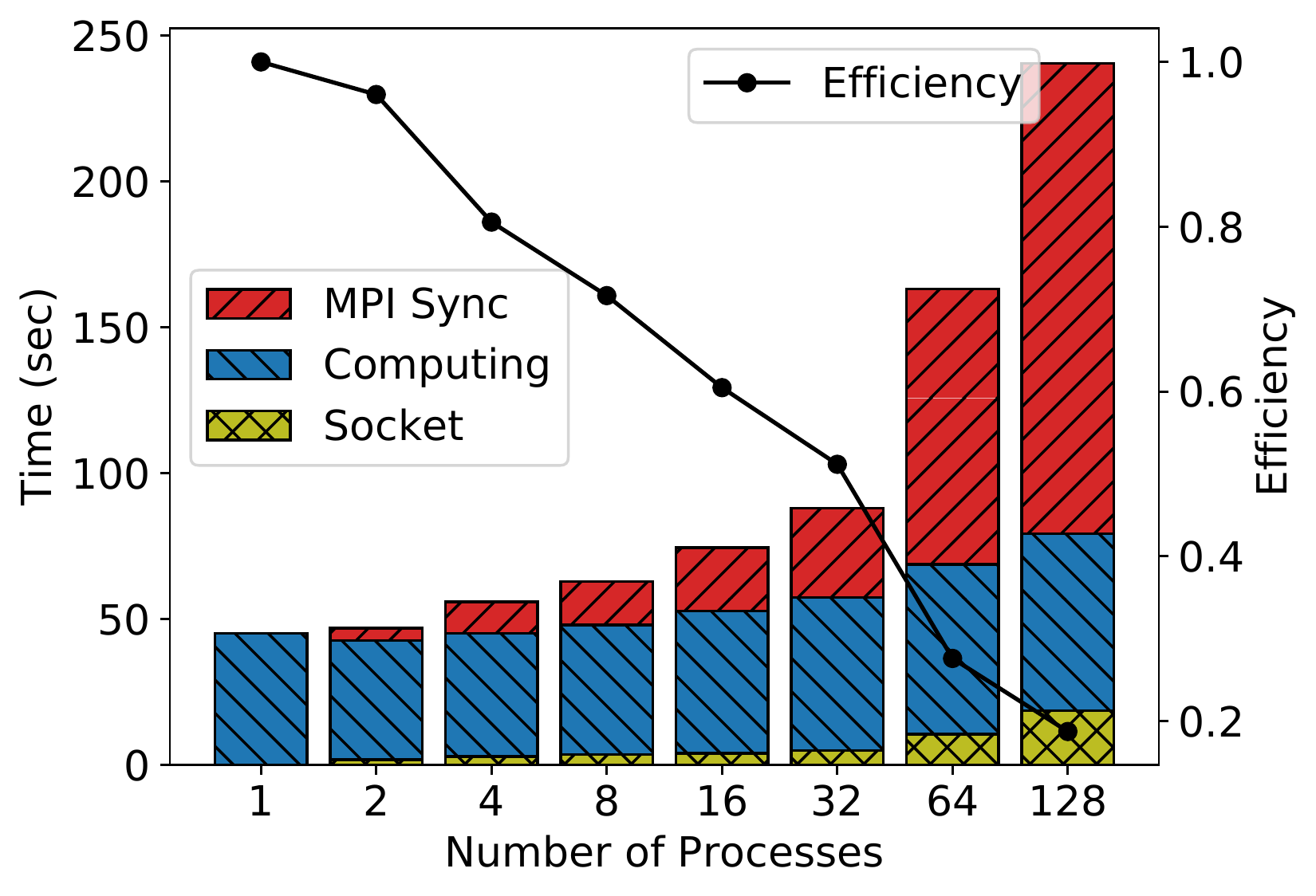}
        \caption{Execution time and efficiency -- weak scaling}
        \label{fig:motivation_weak}
    \end{subfigure}
    \hfill
    \begin{subfigure}[t]{0.25\linewidth}
        \centering
        \includegraphics[width=\linewidth]{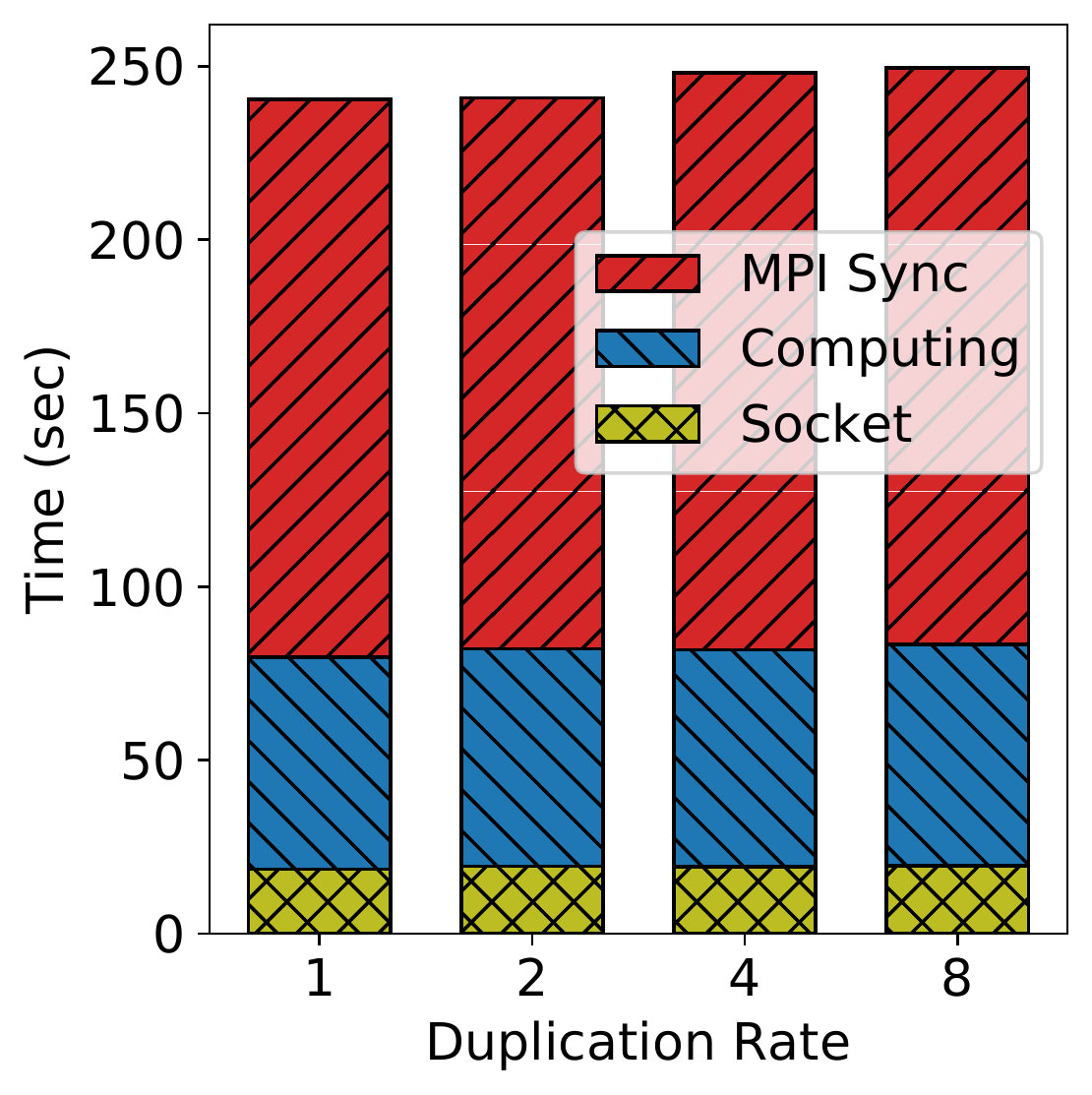}
        \caption{Simulation of 1,024 routers by 128 processes with the duplicated cross-process data}
        \label{fig:duplicate_traffic}
    \end{subfigure}%
    \\
    \begin{subfigure}[t]{0.4\linewidth}
        \centering
        \includegraphics[width=\linewidth]{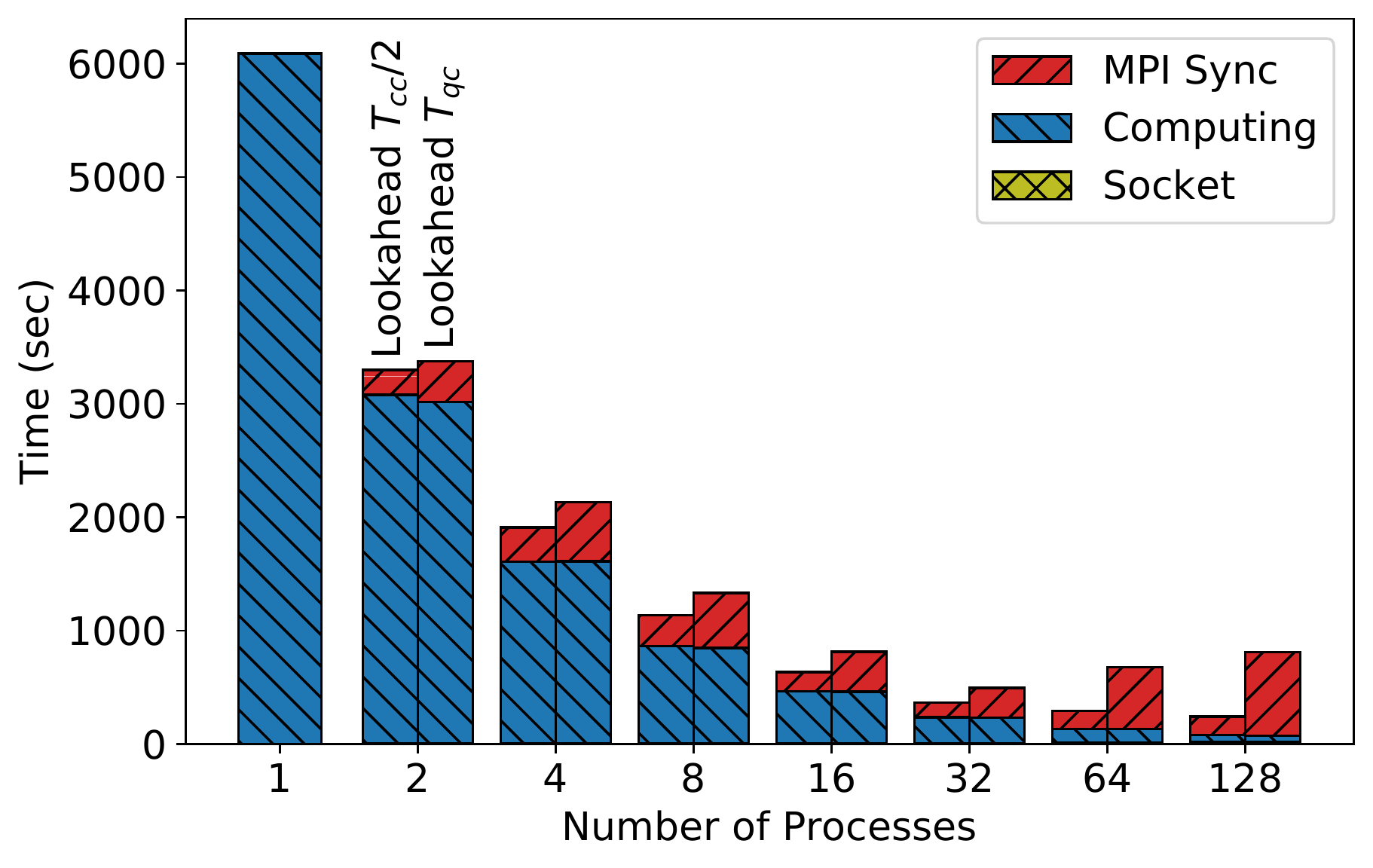}
        \caption{Performance comparison with different lookahead values -- strong scaling}
        \label{fig:lookahead_strong}
    \end{subfigure}%
    \hspace{1em}
    \begin{subfigure}[t]{0.4\linewidth}
        \centering
        \includegraphics[width=\linewidth]{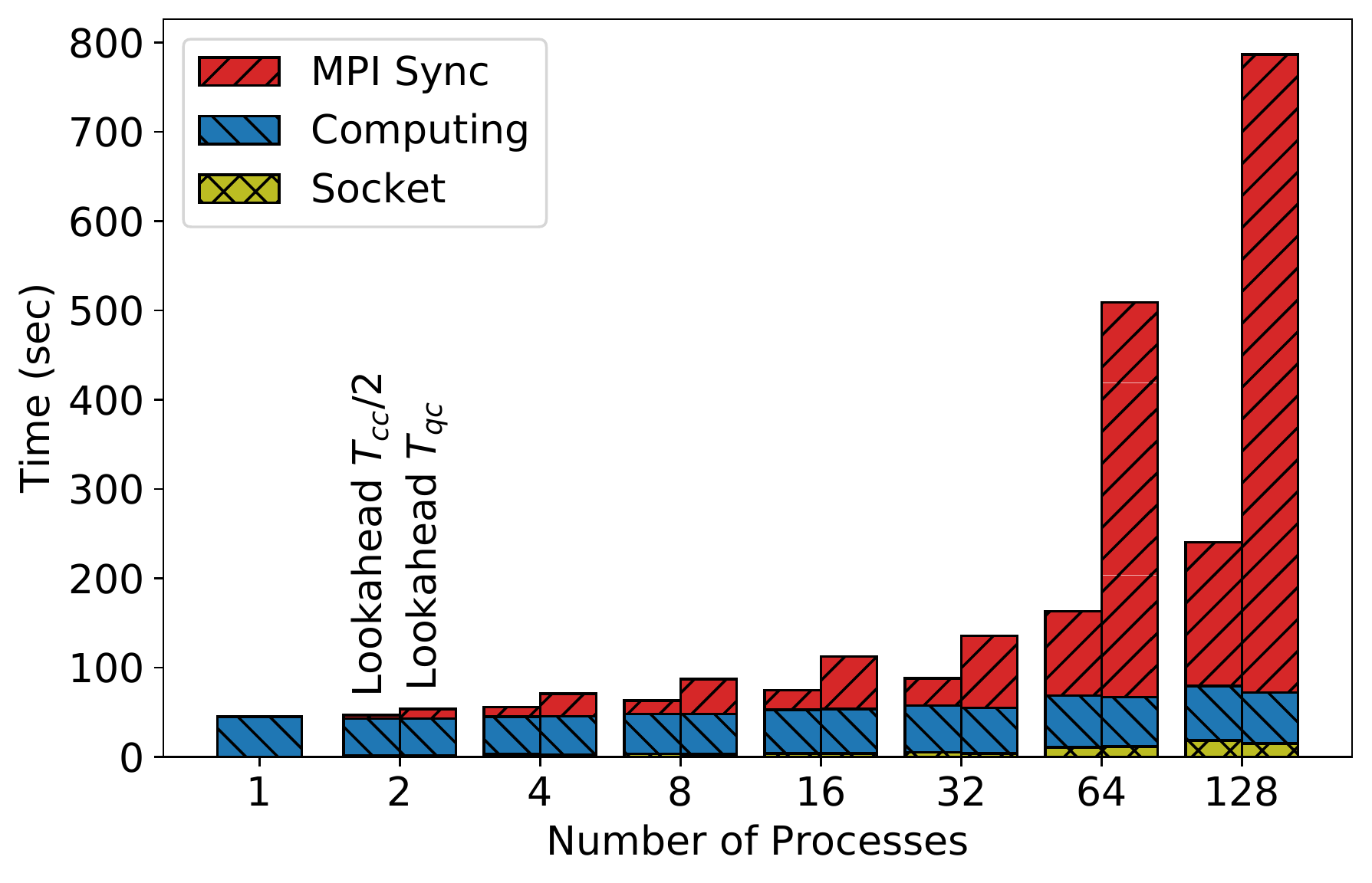}
        \caption{Performance comparison with different lookahead values -- weak scaling}
        \label{fig:lookahead_weak}
    \end{subfigure}
    \caption{Simulations of a linear network}
    \label{fig:linear}
\end{figure}

The first kind of network has a quantum router chain that serves only one flow that distributes entanglement between the two end routers. The quantum network uses a chain of routers to mitigate the exponentially increased loss rate of photons caused by the increased distance~\cite{repeaters}. Figure~\ref{fig:chain_topo} illustrates the topology of a quantum network with five routers. In this type of network, adjacent nodes first use the meet-in-the-middle generation scheme~\cite{generation_schemes} to distribute entanglement over a short distance. Then, the entanglement swapping~\cite{swapping} and purification protocols~\cite{entanglement_purification} are used to distribute entanglement over a longer distance with high fidelity. For the parallel simulation experiments, we evenly divide the number of routers by the number of processes ($n = 1024/p$) and assign them sequentially to each process (i.e., routers 1 to n will be assigned process 1, routers n+1 to 2n to process 2, etc.).
For simulations with the increased lookahead value introduced in Section~\ref{sec:async}, the BSM nodes that connect two different processes are placed in a separate process where the simulation time is allowed to lag.

All our linear topology simulations use a 1 km distance between all adjacent routers. The BSM node is in the middle of each link; in other words,  the length of optical fiber is 0.5 km. All routers contain 100 quantum memories. 

We first evaluate the strong scaling of the simulator in a quantum network with 1,024 routers and a total of 102,400 quantum memories. Strong scaling requires simulations with a fixed workload and an increasing number of processes. We simulate the network with 1 to 128 processes. For this and future simulations using 1 process, the sequential version of SeQUeNCe was used (rather than the parallel version running with 1 MPI process). Since one computing node contains only 36 cores, parallel simulations with 64 and 128 processes are executed on 2 and 4 computing nodes, respectively. Figure~\ref{fig:motivation_strong} illustrates the average execution time of each simulation process for 10 separate simulations and the speedup ratio achieved by the parallel simulation. Since the standard deviation in measured execution time ranges from  7.4 to  189.6 seconds for strong scaling (with 32 and 1 processes, respectively) and from 0.8 to 13.48 seconds for weak scaling (with 16 and 128 processes, respectively), we do not include error bars in the plot. We observe that the execution time decreases with the increasing number of processes. The parallel simulation can speed up the simulation by a factor up to 25. When we double the number of processors, the execution time is halved (indicated by the blue part). However, the synchronization time occupies a larger portion of the total time with the growing number of processes. We also notice that the time for the socket communication between the local and global QSMs is trivial in this case. The time of socket communication is increased from 2.0 seconds to 18.6 seconds as the number of processes increases from 2 to 128. 

We also evaluate the performance of the simulator with weak scaling using a scaled workload. We fix the number of routers on one process to 8 and increase the size of the network with the number of processes. Figure~\ref{fig:motivation_weak} illustrates the execution time and efficiency of the simulations. We observe that the execution time increases with the number of processes. The majority of overhead results from the synchronization algorithm;  little overhead is generated from the socket communication. Local QSMs process more than 90\% of requests without socket communication. Note that we scale the workload by increasing the size of the network; therefore, the workload does not exactly increase linearly and thus causes the increased time of event execution. In addition, the parallelization efficiency decreases with the number of process resulting from the increased overhead. The efficiency of the 2-process simulation is 0.96, whereas the efficiency of the 128-process simulation is 0.19.

Both strong scaling and weak scaling evaluations show an increasing overhead. We observe that the performance in both cases drops significantly when the number of processes increases from 32 to 64. There are two reasons for this drop. First, the performance of the multithreaded global QSM drops when the number of processes is larger than 32. Since we execute the global QSM on one computer that serves all MPI processes, the capability of the global QSM is limited by the number of processors and affected by the MPI processes on the same machine. When the number of processes is less than or equal to 36, the number of processors on one Broadwell node, the global QSM threads do not compete with each other. The block of socket communication on the MPI processes avoids competition between threads and MPI processes. However, when the number of processes is greater than 36, threads and MPI processes compete for computation resources and thus show worse performance. We observe that the global QSM achieves a similar processing speed for parallel simulation with up to 32 processes. The processing time of the global QSM is doubled for simulations with 64 processes and tripled for 128 processes. The degraded performance is also demonstrated with the growing socket communication time as the number of processes increases from 32 to 128 (see Figure~\ref{fig:motivation_weak}).

The second overhead is due to the unbalanced workload. The red bars in Figure~\ref{fig:linear} indicate the synchronization overhead including  the unbalanced workload and  the transmission of data. Therefore, we design further experiments to distinguish and quantify which one actually results in the increased synchronization time.
The experiments duplicate the cross-process data before exchanging events. When the processes receive data from other processes, the duplicated data is discarded. These experiments increase the amount of cross-process transmission during synchronization without affecting the distribution of the workload. Figure~\ref{fig:duplicate_traffic} shows the execution time of simulations that duplicate data from 0 (i.e., no duplication) to 8 times. The four simulations show a similar time in terms of not only event execution and socket communication but also synchronization. Therefore, we conclude that the unbalanced workload during the time window causes a large overhead even if the aggregated workload is balanced. Since the workload balance depends on the simulation setup, addressing this kind of overhead is out of the scope of the design of a parallel simulator.

We also evaluate the lookahead value improvement technique with both strong scaling and weak scaling. Figures~\ref{fig:lookahead_strong} and \ref{fig:lookahead_weak} illustrate the performance of the simulator with and without this technique, respectively. For parallel simulation, the left bar represents the larger lookahead value ($T_{cc}/2$), and the right bar represents the smaller lookahead value ($T_{qc}$). As a result, the lookahead value is increased from 2.5 \si{\micro\second} to 0.15 ms. This proposed technique reduces the simulation synchronization overhead in all cases and thus improves the scalability of the simulator. We also verify that the number of executed simulation events is identical in all cases, which suggests that the increased lookahead time does not affect the simulation accuracy.

\subsection{Quantum Network with Caveman Graph Topology}
\label{subsec:caveman}

\begin{figure}[]
    \centering
    \begin{subfigure}[t]{0.3\linewidth}
        \centering
        \includegraphics[width=\linewidth]{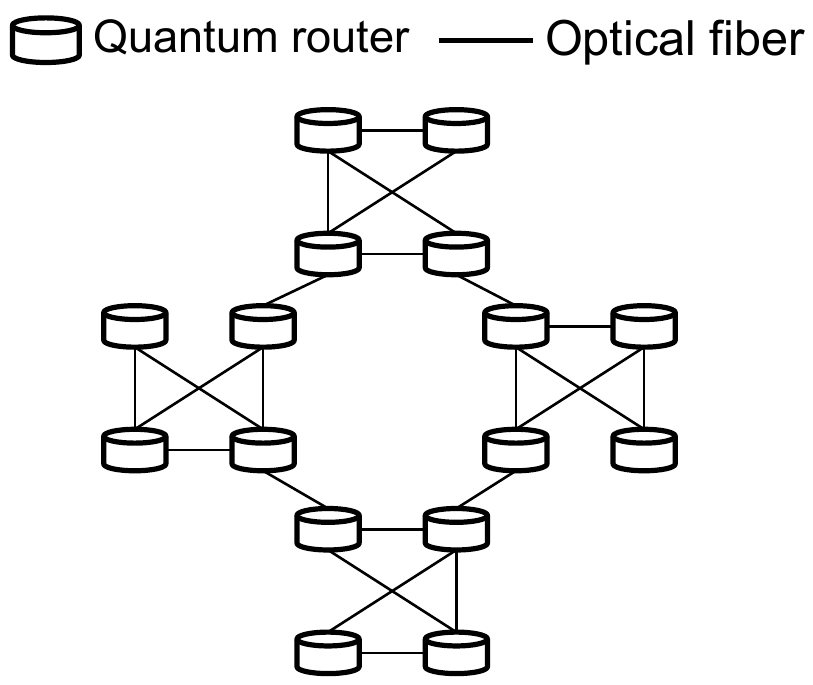}
        \caption{A small-scale network topology example ($Caveman(4,4)$). BSM nodes in the middle of links are not plotted.}
        \label{fig:caveman_topo}
    \end{subfigure}%
    \hspace{1em}
    \centering
    \begin{subfigure}[t]{0.4\linewidth}
        \centering
        \includegraphics[width=\linewidth]{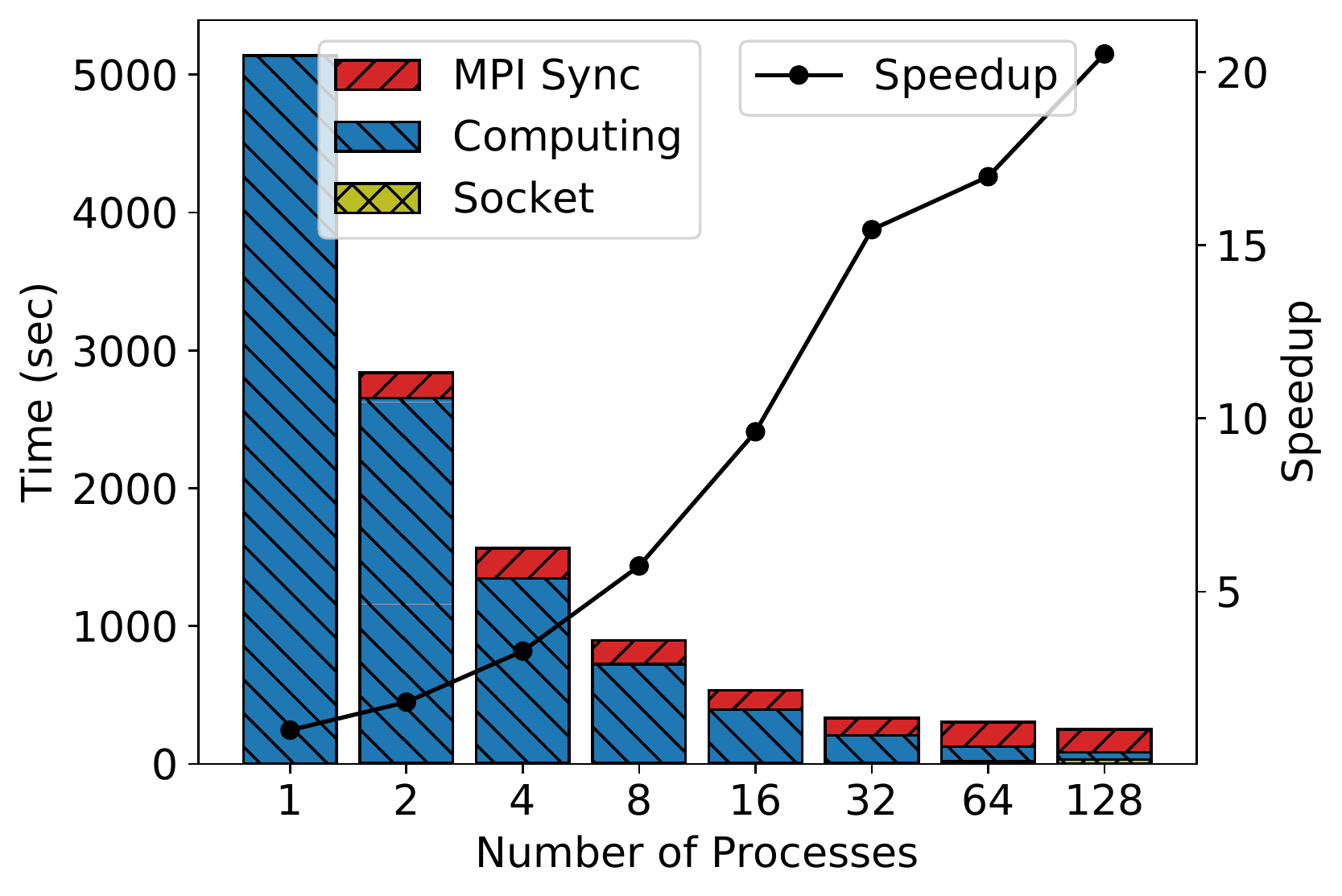}
        \caption{Execution time and speedup for a caveman network with 1,024 nodes in 8 caves}
        \label{fig:caveman_data}
    \end{subfigure}%
    \caption{Simulations of a caveman network}
\end{figure}

The second network uses the caveman graph topology~\cite{caveman} to form the quantum network. The caveman graph $Caveman(n,k)$ is formed from a set of $n$ isolated $k$-node caves by removing one edge from each cave and using it to connect a neighboring cave along a cycle. The community cave structure of the caveman graph allows us to easily group nodes into sets of nodes for parallelization. Nodes in the same cave are assigned to the same process, and neighboring caves are placed together. Compared with networks in Section~\ref{subsec:chain}, this network serves multiple flows with a smaller number of hops. Figure~\ref{fig:caveman_topo} illustrates the network using the caveman graph $Caveman(4,4)$ as the topology, which denotes that both the size of one cave and the number of caves are 4.

All simulations use the $Caveman(128,8)$ network topology that serves 1,024 random flows. Every node in the caveman graph represents one quantum router, and every edge in the graph represents optical fibers and the BSM node used for the meet-in-the-middle protocol. The distance between adjacent routers is 10 km. A sufficient number of quantum memories is allocated on each node to support all flows that pass through them. In our setup 81,500 quantum memories are used in total. At the beginning of the simulation, every router produces a flow that occupies 25 memories in the source and destination routers of the flow and 50 memories in all intermediate routers. The destination routers are selected randomly based on the number of hops in the flow. The number of hops follows an exponential distribution with $\lambda = 1$. We assume that the number of short-distance flows is greater than the number of long-distance flows in a realistic quantum network.

Figure~\ref{fig:caveman_data} illustrates the performance of simulations executed on 1 to 128 processes, with execution time shown as an average of all processes for 5 simulations. Again, we note that the standard deviation in execution time ranges from only 6.0 to 164.0 seconds and thus error bars are not plotted. We evaluate the scalability of the parallel simulation by inspecting the execution time and the speedup with strong scaling. We observe that the execution time decreases as the number of processes grows. The parallel simulation reduces the execution time by a factor up to 20.5. Although the speedup rate of simulating this network is similar to the speedup rate shown in Section~\ref{subsec:chain} in scenarios with fewer than 64 processors, the speedup rate decreases around 23\% for parallel simulations with 64 and 128 processors. The scalability degradation results from an unbalanced workload. Here we use the number of quantum memories utilized on each process to quantify load imbalance on MPI processes. We calculate the coefficient of variation ($CV$), which divides the standard deviation by the mean. We observe that the value of $CV$ increases from 0.01 (2 processes) to 0.17 (128 processes). The unbalanced distribution of quantum memories causes the unbalanced workload, which increases the overhead and therefore reduces the scalability of parallelization. However, placing routers in the same cluster within one process significantly reduces the amount of cross-process entanglement. Less than 10\% of work is forwarded to the global QSM. Even if the 128-process simulation has the most cross-process entanglement states, the simulation only incurs 30-second wait time due to responses from the socket communication. The partition scheme used for caveman networks balances the number of nodes in processes and reduces the number of cross-process entanglement states, but the unbalanced distribution of quantum memories still reduces the scalability. Therefore, we use the next simulation to investigate how partition schemes affect the scalability of parallel simulation on a more complicated topology. 

\subsection{Quantum Network with Autonomous System Network Topology}
\label{subsec:as_net} 

\begin{figure}[]
    \centering
    \begin{subfigure}[t]{0.4\linewidth}
        \centering
        \includegraphics[width=\linewidth]{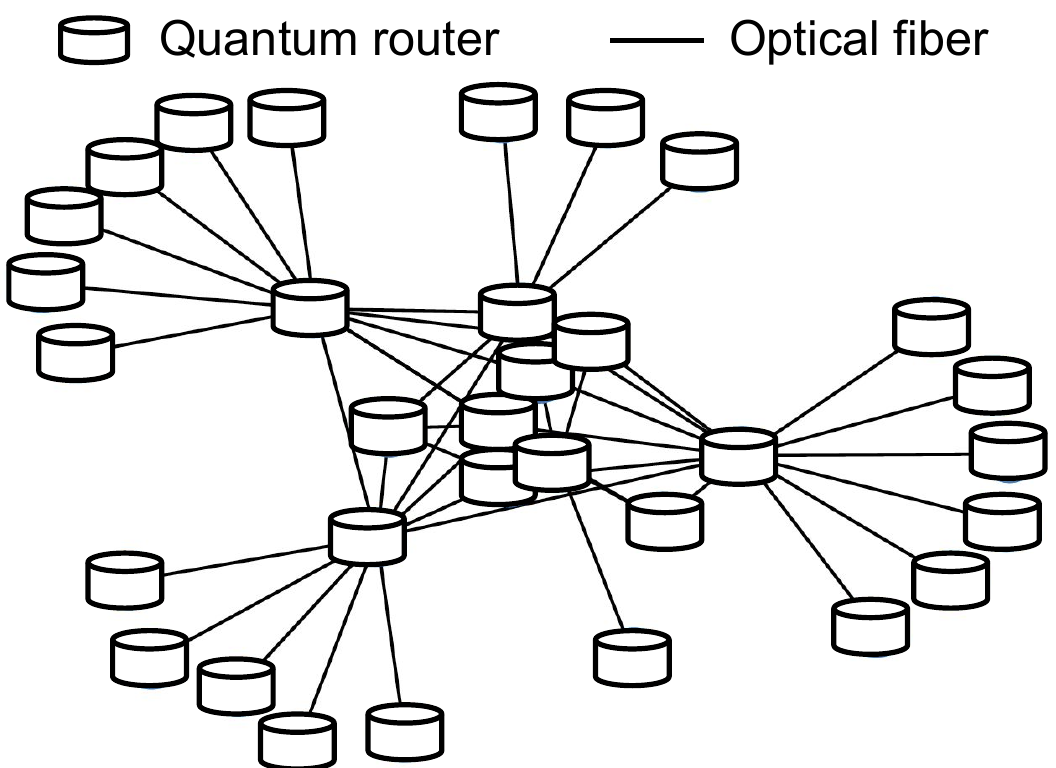}
        \caption{A small-scale network topology example ($AS(32,1)$). BSM nodes in the middle of links are not plotted.}
        \label{fig:as_topo}
    \end{subfigure}%
    \hfill
    \centering
    \begin{subfigure}[t]{0.55\linewidth}
        \centering
        \includegraphics[width=\linewidth]{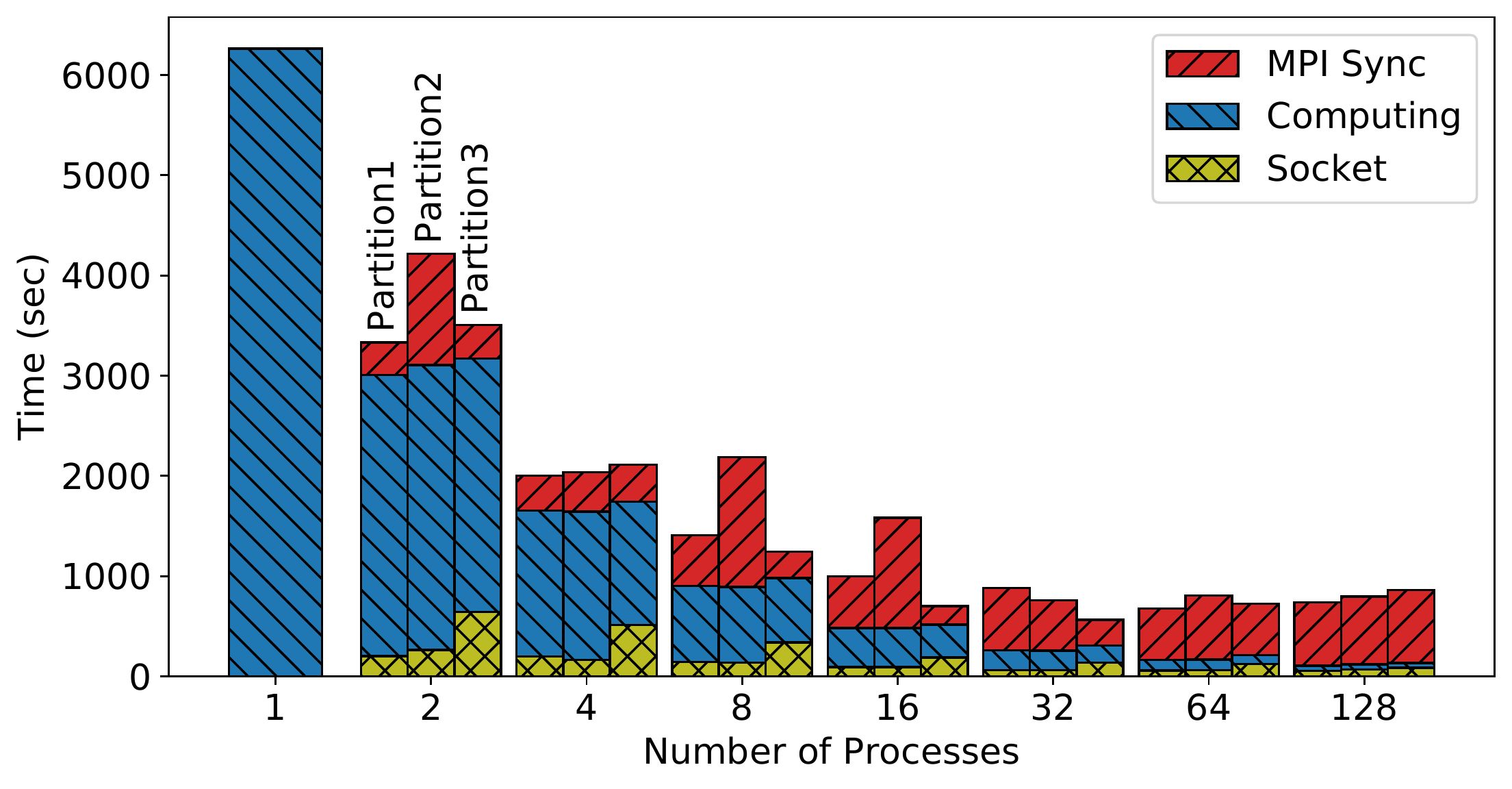}
        \caption{Performance comparison with different partition algorithms}
        \label{fig:as_perf}
    \end{subfigure}%
    \\
    \centering
    \begin{subfigure}[t]{0.5\linewidth}
        \centering
        \includegraphics[width=\linewidth]{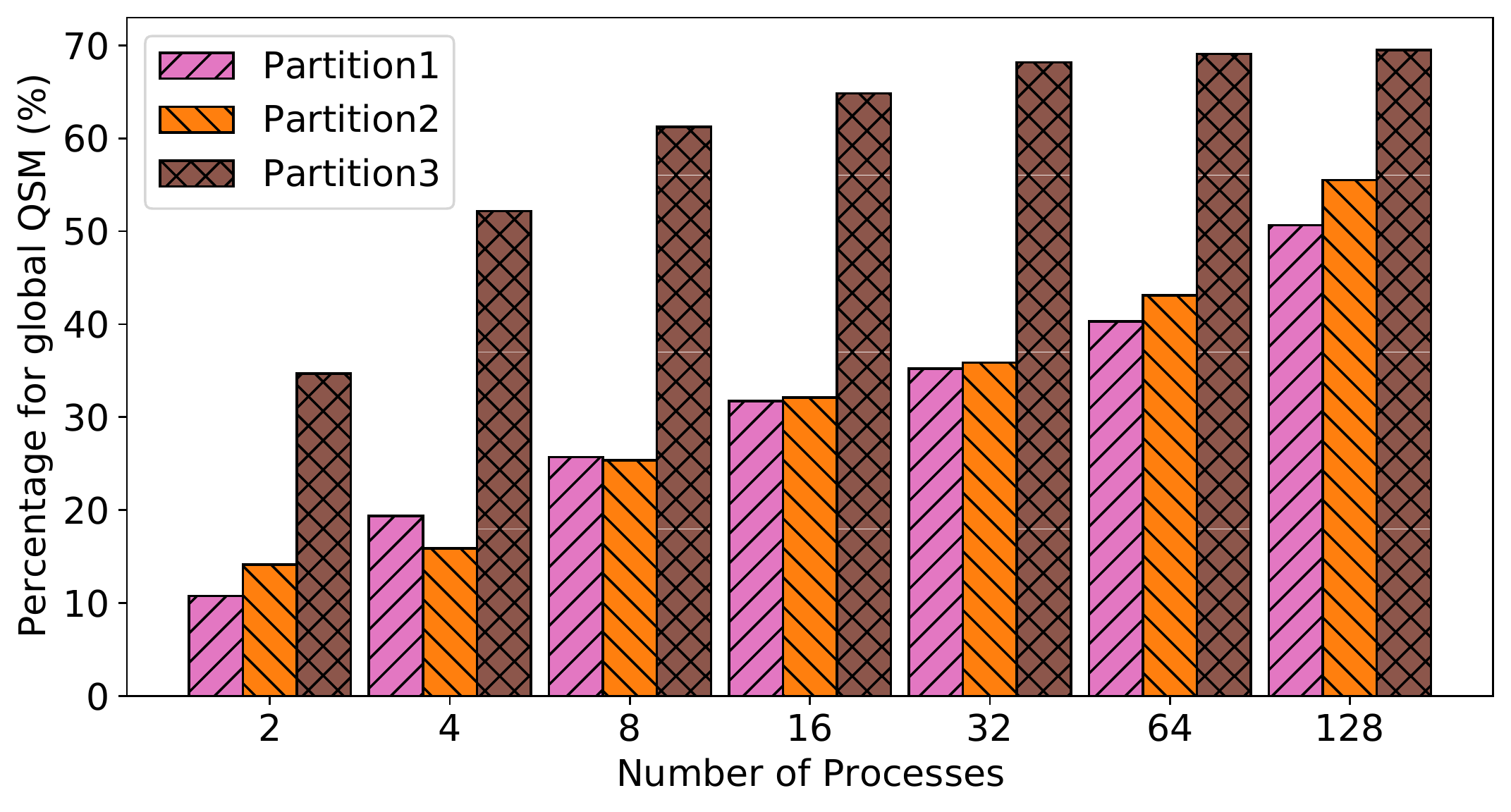}
        \caption{Percentage for requests processed by the global QSM.}
        \label{fig:as_global_percentage}
    \end{subfigure}%
    \centering
    \begin{subfigure}[t]{0.5\linewidth}
        \centering
        \includegraphics[width=\linewidth]{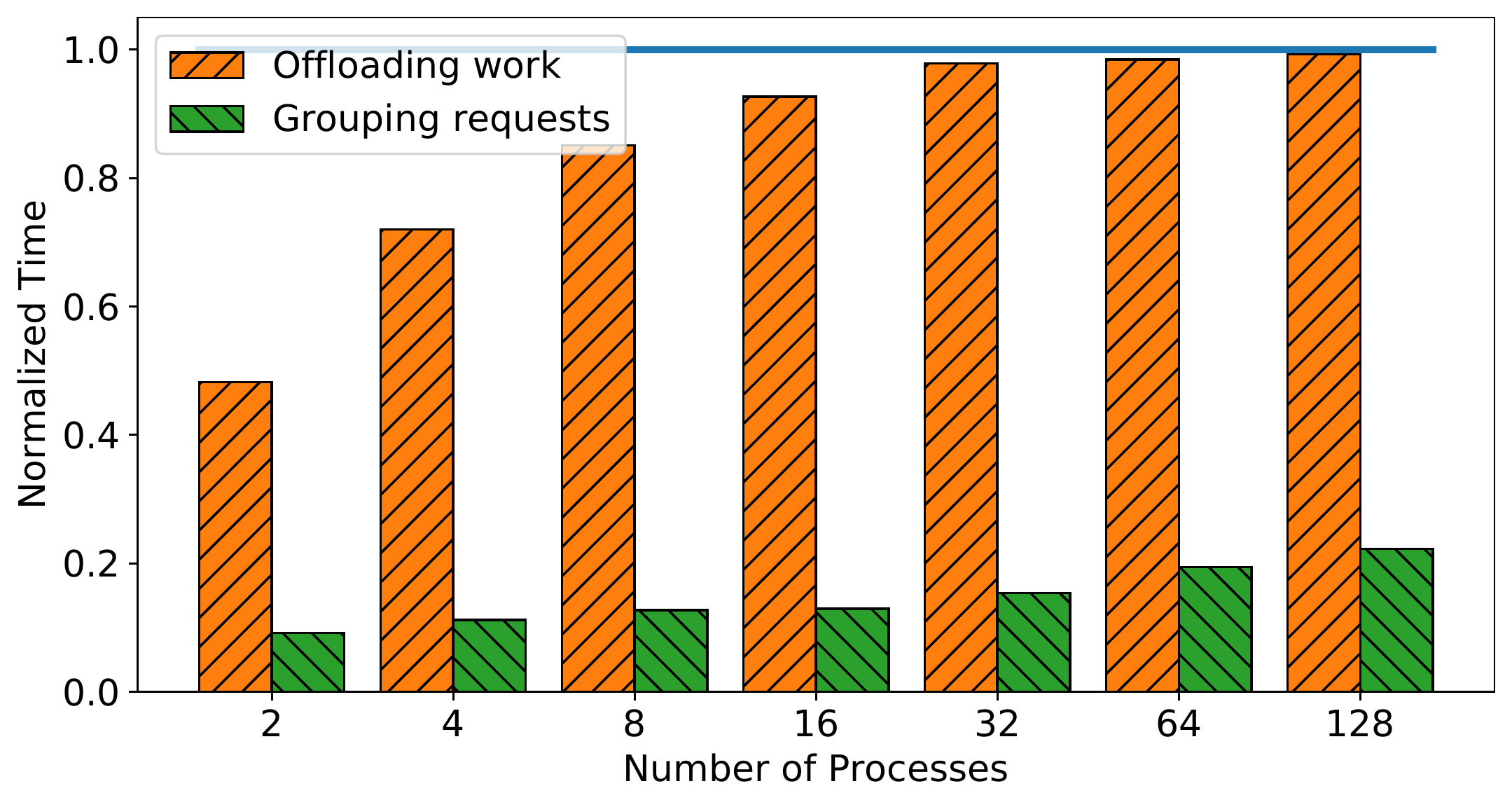}
        \caption{Performance comparison with communication optimization}
        \label{fig:as_optimizaton}
    \end{subfigure}%
    \caption{Simulations of an autonomous system network graph}
\end{figure}
 
We choose the graph resembling the Internet autonomous system (AS) network as the topology of the third simulated network. As with previous simulations, nodes of the graph represent quantum routers, and edges represent optical fibers with BSM nodes used for the meet-in-the-middle protocol. These routers follow the same method introduced in Section~\ref{subsec:caveman} to produce random flows. We simulate the quantum network with 1,024 routers in the AS network topology. There are 81,300 quantum memories used (again, based on flow usage) to serve 1,024 random flows. We investigate the scalability of the parallel simulator under this scenario. We use Networkx~\cite{networkx} to generate the graph $AS(1024,0)$ that denotes a random 1024-node graph using seed 0. 

Figure~\ref{fig:as_topo} illustrates a smaller network $AS(32,1)$. Unlike the previous two network topologies, this network topology includes many edge routers that are connected by the core routers. These core routers have a higher degree than the edge routers, which results in an unbalanced distribution of workload among the routers. The unbalanced workload on the routers is a challenging scenario for the network partition algorithm for parallel simulation.

We propose three heuristic solutions for network partitioning: $Partition 1$, $Partition 2$, and $Partition 3$. The proposed solutions use simulated annealing~\cite{simulated_anneal} to optimize their energy functions. We define three required components for the simulated annealing: \textbf{state}, \textbf{neighboring state}, and \textbf{energy function}. \textbf{State} defines the mapping relation between nodes and processes. The initial state evenly assigns routers to processes. \textbf{Neighboring state} is the state that could be generated by swapping two routers in different processes. The initial state and neighboring state make the partition scheme always distribute nodes to processes evenly. The simulated annealing method starts from the initial state and randomly jumps between neighboring states. After a few iterations, the state with the lowest energy is the final partition scheme. Given different energy functions, simulated annealing may produce different partition schemes. We design three \textbf{energy functions} using different assumptions.

\begin{enumerate}
    \item $Partition 1$ assumes that users know all flows in the simulation. The energy function is to calculate the number of cross-process flows. 
    \item $Partition 2$ assumes that users know only the topology of the network. The energy function is to calculate the number of cross-process quantum channels.
    \item $Partition 3$ assumes that users know how many memories are used in each router. The energy function is designed to calculate the imbalance of memory distribution. 
\end{enumerate}

The energy functions of $Partition 1$ and $Partition 2$ are designed to reduce the overhead of the socket communication, while the energy function of $Partition 3$ is designed to reduce the overhead of unbalanced workloads. We conduct parallel simulations with these three partition solutions on 2 to 128 processes. These simulations help us explore how various network partition affects the parallelization performance.

Figure~\ref{fig:as_perf} illustrates the execution time of the simulations. We note again that the maximum standard deviation in simulation time is  220.8 seconds, and so error bars are not plotted. For parallel simulations, the left, middle, and right bars represent partitioning using $Partition 1$, $Partition 2$, and $Partition 3$, respectively. We observe that parallel simulations using $Partition 3$ consume more time for socket communication than the other two partition methods because of the different objective functions. Figure~\ref{fig:as_global_percentage} demonstrates that the increased socket communication time results from the additional requests processed by the global QSM. Compared with $Partition 1$, $Partition 3$ requires the local QSM to forward around 30\% requests to the global QSM.  Among the three methods, $Partition 1$ performs the best when the number of processes is 2, 4, 64, and 128; $Partition 2$ performs the worst in most cases; $Partition 3$ performs the best when the number of processes is 8, 16, and 32. When the parallel simulation uses fewer processes, $Partition 1$ finds a partition with low overhead from the socket communication, and the imbalance of workload is suppressed by the limited number of processes as well as the equal number of nodes in the processes. As the number of processes increases, the overhead from the unbalanced workload dominates the execution time, and the socket communication time is mitigated by the multithreaded global QSM. Thus, $Partition 3$ shows the best parallel performance. However, when the number of processes is too large, equally distributing nodes to processes becomes a bottleneck in $Partition 3$. For example, with a parallel simulation of 128 processes, we expect every process to control $81300 / 128 \approx 89$ memories on average, but the heaviest router controls 2,675 memories. Seven routers need to be placed in the same process with the heaviest router, which deteriorates the workload balance. In addition to the decreased overhead on socket communication, three partition solutions show similar performance when the number of processes is too large. 

$Partition 3$ demonstrates the largest socket communication time. Therefore, those scenarios are good candidates for evaluating the proposed techniques to reduce communication time, as described in Section~\ref{subsec:server}. We compare the performance of the parallel simulation with work offloading and requests grouping, respectively. Figure~\ref{fig:as_optimizaton} illustrates the normalized execution time of the simulation. We normalize the execution time of parallel simulations by dividing the execution time of the parallel simulation that ignores optimizations on the socket communication. We observe that both optimization techniques always speed up simulations. Offloading work from the global QSM to the local QSMs achieves better performance for simulations with fewer processes. Since the simulation using fewer processes leads to less cross-process entanglement, more manipulations of quantum states can be handled locally instead of using the global QSM. Compared with work offloading, grouping multiple requests to one message achieves a more significant performance improvement, especially for simulations with more processes. By grouping requests, parallel simulations can save more than 75\% of the communication time. Moreover, the benefit from this optimization takes less impact from the increased number of processes. After aggregating both optimization methods, we speed up the simulation using 128 processes by 4.5 times and the simulation using 2 processes by 11.5 times, which significantly improves the scalability of our simulator.

%% file: related_work.tex
\section{Related Work}
\label{sec:related}

Recently, several discrete-event quantum network simulators have been developed, including QuISP~\cite{matsuo2019quantum}, NetSquid~\cite{netsquid}, SQUANCH~\cite{squanch}, and SeQUeNCe~\cite{sequence, wu2019simulations, spw_19}. These simulators offer a flexible and reproducible testing environment to study the performance of quantum networks, but differ in the supported features, hardware and protocol models, as well as simulation use cases. With the exception of SQUANCH, the performance of these simulators has been limited due to their sequential execution of simulation events. We believe that the techniques that allowed us to parallelize quantum state management in the SeQUeNCe simulator can be transferred to the other simulators.

Thus far, large-scale quantum network simulations used sequential processing and achieved improved scalability by relying on simplified qubit representations. For example, QuISP~\cite{matsuo2019quantum} aims to simulate a full Quantum Internet consisting of up to 100 networks of up to 100 nodes each, and scalability is achieved by replacing full quantum-state tracking by maintaining an error model of qubits in large networks. NetSquid allows the user to choose between several quantum state representations that vary in their scalability, namely ket vector, density matrix, stabiliser, and graph state representation. The NetSquid simulator was successfully used in a recent simulation of a chain of up to 1,000 nodes~\cite{netsquid}. SQUANCH~\cite{squanch} is notable for allowing parallelized simulations that use multiprocessing where each process manages one simulated network node. However, the simulator does not attempt to model interactions of protocols or details at the physical layer, and it does not guarantee the order of event executions.


Several classical network simulators support parallelization, such as ns-3~\cite{ns3} and OMNeT++~\cite{OMNET}. However, parallelization of classical network simulators is simpler due to the absence of exponentially large quantum state representations and non-local effects of quantum state measurements. Kazer et al.~\cite{kazer2018fast} used the built-in MPI-based PDES framework of OMNeT++ to simulate a highly interconnected data center network and found that synchronization can actually cause parallel DES to perform worse than a single-threaded implementation. Nikolaev et al.~\cite{ns3-perf} evaluated the performance of the distributed ns-3 network simulator and concluded that it scales only to about 100--200 cores/processes. Beyond that limit, performance suffers because of increased interprocess synchronization and communication costs. Barnes et al.~\cite{barnes2011benchmark,billion-nodes-ns3,ns3-perf} simulated networks with 1 billion nodes~(routers and hosts) using ns-3 on a supercomputer. Their scalability metric and their implementation of a NULL message synchronization algorithm~\cite{pdes} may be useful in future work.



Our recent work~\cite{parallel_qkd} demonstrated that parallelization can greatly speed up the simulation of quantum key distribution networks~\cite{jin2019genuine}. In that work, a network model was effectively divided into multiple components, each of which was executed on a different logic process with a synchronization mechanism to ensure the correct execution order of cross-process events. A QKD network composed of QKD terminals and channels for transmitting photons as well as packets was simulated. Similarly as in this work, we applied the conservative synchronization algorithm~\cite{conservative}. However, this earlier work was limited to tracing quantum superposition states and did not consider distribution of entanglement which is necessary in general-purpose quantum communication networks. In this paper we address this shortcoming by designing and implementing simulation techniques that use distributed quantum state management that allows efficient parallel simulation of entanglement distribution.

%% file: conclusion.tex
\section{Conclusion} \label{sec:conclusion}

In this paper we present a parallelized version of the SeQUeNCe simulator for quantum communication networks. The simulator uses a conservative synchronization algorithm for maintaining event execution order between parallel processes. A quantum state manager (QSM) was developed to track shared quantum states among processes, with a centralized global QSM servicing requests from local QSMs on individual processes. We additionally investigate multiple methods for optimizing the performance of a quantum network simulator.

Furthermore, by working within the modularized architecture of the SeQUeNCe simulator, we provide a parallel simulator that requires minimal adaptations to sequentially run simulations. The kernel module, which processes events and stores quantum states, was fully parallelized while leaving the higher-level hardware and protocol layers with few changes. Our design ensures identical results for sequential and parallel simulations in conjunction with the well-ordered event execution guaranteed by the conservative synchronization algorithm.

We are releasing the parallelized SeQUeNCe simulator as open-source software with the goal of providing an easy avenue for constructing parallel simulations whose performance will enable development of future large-scale quantum networks. Our plans for future development of the SeQUeNCe simulator include introduction of new optical hardware models, including new types of quantum memories, and the release of an interactive GUI for the intuitive construction of new simulations.